\documentclass[preprint2]{emulateapj-rtx4} 
\usepackage{amsmath,natbib,graphicx}
\usepackage{epsf}
\input{epsf}
\usepackage{epstopdf}
\usepackage{multirow}
\usepackage{epsfig}
\usepackage{color}
\usepackage{lscape}
\DeclareGraphicsExtensions{.jpg,.pdf,.png,.eps,.ps}

\DeclareGraphicsExtensions{.jpg,.pdf,.png,.eps,.ps}

\newcommand{\ltsima}{$\; \buildrel < \over \sim \;$}
\newcommand{\ltsim}{\lower.5ex\hbox{\ltsima}}

\newcommand{\be}{\begin{equation}}
\newcommand{\ee}{\end{equation}}
\newcommand{\bea}{\begin{eqnarray}}
\newcommand{\eea}{\end{eqnarray}}

\begin{document}

\title{Angular Power Spectra of the Millimeter-Wavelength Background Light from Dusty Star-forming Galaxies with the South Pole Telescope}

\author{N.~R.~Hall,\altaffilmark{1}
R.~Keisler,\altaffilmark{5,7}
L.~Knox,\altaffilmark{1} C.~L.~Reichardt,\altaffilmark{2}
P.~A.~R.~Ade,\altaffilmark{3} K.~A.~Aird,\altaffilmark{4}
B.~A.~Benson,\altaffilmark{2,5,6} L.~E.~Bleem,\altaffilmark{5,7} J.~E.~Carlstrom,\altaffilmark{5,6,7,8} 
C.~L.~Chang,\altaffilmark{5,6} 
H.-M. Cho,\altaffilmark{2}  
T.~M.~Crawford,\altaffilmark{5,8} 
A.~T.~Crites,\altaffilmark{5,8} 
T.~de~Haan,\altaffilmark{9} 
M.~A.~Dobbs,\altaffilmark{9}
E.~M.~George,\altaffilmark{2}
N.~W.~Halverson,\altaffilmark{10} 
G.~P.~Holder,\altaffilmark{9}
W.~L.~Holzapfel,\altaffilmark{2} J.~D.~Hrubes,\altaffilmark{4}
M.~Joy,\altaffilmark{11}
A.~T.~Lee,\altaffilmark{2,12} E.~M.~Leitch,\altaffilmark{5,8}
M.~Lueker,\altaffilmark{2}
J.~J.~McMahon,\altaffilmark{5,6} J.~Mehl,\altaffilmark{2}
S.~S.~Meyer,\altaffilmark{5,6,7,8} J.~J.~Mohr,\altaffilmark{13,14,15}
T.~E.~Montroy,\altaffilmark{16} 
S.~Padin,\altaffilmark{5,8} T.~Plagge,\altaffilmark{2}
C.~Pryke,\altaffilmark{5,6,8} 
J.~E.~Ruhl,\altaffilmark{16} K.~K.~Schaffer,\altaffilmark{5,6}
L.~Shaw,\altaffilmark{9} E.~Shirokoff,\altaffilmark{2} 
H.~G.~Spieler,\altaffilmark{12}
B.~Stalder,\altaffilmark{17} 
Z.~Staniszewski,\altaffilmark{16} A.~A.~Stark,\altaffilmark{17} 
E.~R.~Switzer,\altaffilmark{5,6} K.~Vanderlinde,\altaffilmark{9}
J.~D.~Vieira,\altaffilmark{5,7} 
R.~Williamson\altaffilmark{5,8} 
and O.~Zahn\altaffilmark{18} 
}

\altaffiltext{1}{Department of Physics, University of California, One Shields Avenue, Davis, CA 95616, USA}
\altaffiltext{2}{Department of Physics, University of California, Berkeley, CA 94720, USA}
\altaffiltext{3}{Department of Physics and Astronomy, Cardiff University, CF24 3YB, UK}
\altaffiltext{4}{University of Chicago, 5640 South Ellis Avenue, Chicago, IL 60637, USA}
\altaffiltext{5}{Kavli Institute for Cosmological Physics, University of Chicago, 5640 South Ellis Avenue, Chicago, IL 60637, USA}
\altaffiltext{6}{Enrico Fermi Institute, University of Chicago, 5640 South Ellis Avenue, Chicago, IL 60637, USA}
\altaffiltext{7}{Department of Physics, University of Chicago, 5640 South Ellis Avenue, Chicago, IL 60637, USA}
\altaffiltext{8}{Department of Astronomy and Astrophysics, University of Chicago, 5640 South Ellis Avenue, Chicago, IL 60637, USA}
\altaffiltext{9}{Department of Physics, McGill University, 3600 Rue University, Montreal, Quebec H3A 2T8, Canada}
\altaffiltext{10}{Department of Astrophysical and Planetary Sciences and Department of Physics, University of Colorado, Boulder, CO 80309, USA}
\altaffiltext{11}{Department of Space Science, VP62, NASA Marshall Space Flight Center, Huntsville, AL 35812, USA}
\altaffiltext{12}{Physics Division, Lawrence Berkeley National Laboratory, Berkeley, CA 94720, USA}
\altaffiltext{13}{Department of Physics, Ludwig-Maximilians-Universit\"{a}t, Scheinerstr.\ 1, 81679 M\"{u}nchen, Germany}
\altaffiltext{14}{Excellence Cluster Universe, Boltzmannstr.\ 2, 85748 Garching, Germany}
\altaffiltext{15}{Max-Planck-Institut f\"{u}r extraterrestrische Physik, Giessenbachstr.\ 85748 Garching, Germany}
\altaffiltext{16}{Physics Department, Center for Education and Research in Cosmology and Astrophysics, Case Western Reserve University, Cleveland, OH 44106, USA}
\altaffiltext{17}{Harvard-Smithsonian Center for Astrophysics, 60 Garden Street, Cambridge, MA 02138, USA}
\altaffiltext{18}{Berkeley Center for Cosmological Physics, Department of Physics, University of California, and Lawrence Berkeley National Laboratory, Berkeley, CA 94720, USA}

\email{nrhall@ucdavis.edu}

\begin{abstract}
We use data from the first 100 deg$^2$ field observed by the South Pole Telescope (SPT) in 2008 to measure the angular power spectrum of temperature anisotropies contributed by the background of dusty star-forming galaxies (DSFGs) at millimeter wavelengths.  From the auto and cross-correlation of 150 and $220\,$GHz SPT maps, we significantly detect both Poisson distributed and, for the first time at millimeter wavelengths, clustered components of power from a background of DSFGs.  The spectral indices of the Poisson and clustered components are found to be $\bar \alpha_{150-220}^P=3.86\pm0.23$ and $\alpha_{150-220}^C=3.8\pm1.3$, implying a steep scaling of the dust emissivity index $\beta \sim 2$.  The Poisson and clustered power detected in SPT, BLAST (at 600, 860, and 1200$\,$ GHz), and {\it Spitzer} (1900$\,$GHz) data can be understood in the context of a simple model in which all galaxies have the same graybody spectrum with dust emissivity index of $\beta=2$ and dust temperature $T_d=34\,$K.  In this model, half of the $150\,$GHz background light comes from redshifts greater than $3.2$.  We also use the SPT data to place an upper limit on the amplitude of the kinetic Sunyaev--Zel'dovich power spectrum at $\ell = 3000$ of  $13\,\mu {\rm K}^2$ at 95\% confidence.
\end{abstract}


\bigskip\bigskip

\section{Introduction}
\label{sec:intro}

The cosmic infrared background (CIB) is produced by thermal emission
from dust in galaxies over a very broad range in redshift \citep{lagache05,marsden09}.
The dust grains, ranging in size from a few molecules to 0.1 mm,
absorb light at wavelengths smaller than their size, and re-radiate it at longer wavelengths.  
Sufficient absorption occurs to account for roughly equal amounts of energy in
the CIB and in the unprocessed starlight that makes up the optical/UV background \citep{dwek98,fixsen98}.

The spectral shape of the graybody emission from the dust, steeply rising with increasing frequency to a peak near $\sim 3000\,$GHz ($100\, \mu$m), results in a very peculiar luminosity-redshift relation for the galaxies contributing to the CIB (e.g., \citealp{blain02}).  For example, an object with emission that peaks at a typical rest-frame frequency of $3000\,$GHz, has an observed $\nu = 220\,$GHz ($\lambda = 1.4$~mm) flux that is nearly independent of redshift for $1 \la z \la 13$, as the increased luminosity due to increased rest-frame frequency compensates for 
the cosmological dimming with redshift.
Since we do not expect significant emission from redshifts anywhere near $z \gtrsim 13$, measurements of the CIB at $220\,$GHz are sensitive to the complete history of emission from dusty star-forming galaxies (DSFGs).

\citet{puget96} detected the CIB for the first time with data from the Far Infrared Absolute Spectrophotometer (FIRAS) on the {\it Cosmic Background Explorer} ({\it COBE}) satellite.  
The CIB was first partially resolved into light from individual galaxies in
 350$\,$GHz (850$\,\mu$m) surveys using the SCUBA camera on the James Clerk Maxwell Telescope \citep{smail97,hughes98c,eales99}.  
 \citet{chapman05} found that sources with 350$\,$GHz fluxes in the $2\,$--$15\,$mJy range have a mean redshift of $z \simeq 2.2$ and a maximum of $z =  3.6$.  

Progress is being made in understanding the nature of the sources of the CIB.  Using deep {\it Spitzer} data, \citet{dole06} stacked 4.3$\,$THz (70 $\mu$m) and 1.9$\,$THz (160 $\mu$m) flux from more than 19000 12.5$\,$THz (24 $\mu$m) sources to determine that these sources contribute more than 70\% of the CIB at 4.3$\,$THz and 1.9$\,$THz.  \citet{marsden09} perform a similar analysis with Balloon-borne Large-Aperture Submillimeter Telescope (BLAST) maps to show that Far-Infrared Deep Extragalactic Legacy (FIDEL) catalog sources are contributing $\sim95$\% of the FIRAS-determined background flux at $600\,$GHz (500~$\mu$m).  \citet{lefloch05} and \citet{lagache05} show that the fraction of the CIB originating from sources at high redshift decreases with increasing frequency.  \citet{marsden09} confirm this and find that 60\% of the CIB originates from $z>1.2$ sources at $600\,$GHz.  By including sources detected with the assistance of gravitational lensing magnification, it is possible to account for close to 75\% of the 350GHz background \citep{blain99,knudsen08}.  Significant advances in identifying the sources responsible for the CIB are expected with {\it Herschel}/SPIRE's 600, 860, and 1200$\,$GHz channels and their resulting catalogs.  The full-sky surveys of the {\it Planck}/HFI will complement {\it Herschel} by fully exploring the bright and rare end of the flux distribution.

Observations of the diffuse background complement those of individual sources for two reasons.  
First, due to limitations on the resolution of current instruments, the bulk of the light at frequencies lower than 12.5$\,$THz is emitted by sources at or below the confusion limit.\footnote{Usually defined as the flux at which sources reach a surface density of 1 object per 40 beams \citep{condon74}.} 
Second, the clustering properties of the CIB are sensitive to the relation between dust-enshrouded star-forming galaxies and the dark-matter halos in which they reside \citep{haiman00}. 
This connection is critical to any theoretical understanding of the history of star formation.  
Clustering information can also come from very large source catalogs,
 but the number of sources required to compete with diffuse clustering
 measurements is in the tens of thousands \citep{knox01}.

Predictions for the clustered CIB fluctuations date back to the pioneering work of \citet{bond86} and \citet{bond91b}.  
Interest was revived with the SCUBA measurements, the discovery and characterization of the background from {\it COBE} data \citep{puget96,fixsen98}, and the prospect of far-IR to millimeter wavelength sky maps to come from a number of cosmic microwave background (CMB) mapping projects, including {\it Planck} \citep{planck06}.  
Predictions for the clustered and Poisson power of ``unresolved SCUBA sources'' have been made by a number of authors \citep{scott99,haiman00,knox01,magliocchetti01,perrotta03,song03,negrello07,fernandez-conde08,righi08}.

Models generically predict a clustering amplitude on the order of a few percent of the intensity of the mean background and angular scales of a few arcminutes, which can be understood by the following argument.  A few arcminutes corresponds to several comoving Mpc at cosmological distances.  The variance in density fluctuations is of order unity on Mpc scales, and the fluctuations in the galaxy distribution will be larger by some bias factor.  The integrated fluctuations along the line of sight probe of order a few hundred independent cells; the fluctuations are thus suppressed by the square root of a few hundred.  This leads to an expected clustering amplitude on the order of a few percent of the mean.

The first detection of diffuse infrared background clustered power came from 160$\,\mu$m {\it Spitzer} data
\citep{grossan07,lagache07}.  This was followed recently by BLAST which measured the clustered power near the energetic peak of the CIB at 600, 860, and 1200$\,$GHz (500, 350, and $250\,\mu$m) \citep{viero09}.  
Here,  we report on the first detection of CIB clustered power at millimeter wavelengths, and compare models of the CIB with measurements of this clustered power over more than a decade in frequency.  

This paper is one of a set of companion papers presenting early results from South Pole Telescope (SPT) observations that include \citet[hereafter V09]{vieira09} and \citet[hereafter L10]{lueker09}. 
V09 present detections and spectral index measurements of point sources that are sufficiently bright to be resolved from the diffuse background.
L10 present fine scale CMB anisotropy bandpowers measured with SPT data and report the detection of the thermal Sunyaev--Zel'dovich (tSZ) power spectrum in a ``DSFG-subtracted'' linear combination of the 150 and $220\,$GHz maps.   
In support of the interpretation of the L10 power spectrum, this paper constrains the amount of residual CIB clustered power that can remain in that map despite the DSFG subtraction.

To interpret our clustering signal as arising from the clustering of DSFGs, we need to consider other possible sources of fluctuation power as well.  We address the impact on our results of the predicted Poisson power from radio galaxies.  We also use a number of lines of argument to conclude that the contributions from the clustering of radio galaxies are negligible for both the 150 and $220\,$GHz maps.  

A difficult contaminant for the interpretation of our data is the kinetic Sunyaev-Zel'dovich (kSZ) effect, which has a highly uncertain amplitude.  We discuss implications of the kSZ component for our determination of the clustering signal, and also provide an upper limit on kSZ power sufficiently tight to rule out the highest predictions in the literature.  

The paper is organized as follows.  We describe the external galaxy emission models we use in Section 2.  
In Section 3, we describe our modeling of other signals, namely, the CMB, tSZ, kSZ, and Galactic cirrus emission.  In Section 4, we describe the data and its processing to $150\,$GHz auto spectrum bandpowers (referred to by $150\times150$), the 150 and $220\,$GHz cross-spectrum ($150\times220$), and $220\,$GHz auto spectrum bandpowers ($220\times220$). In Section 5, we fit the data to the power spectrum templates described in Sections 2 and 3.  
In Section 6, we discuss the implications of these results when combined with those of other experiments and compare the results with models for the properties of the sources composing the millimeter-wavelength to far-infrared background light.  In Section 7, we calculate the expected level of residual contamination from \nolinebreak{DSFGs} in the DSFG-subtracted map of L10.  We summarize the main results in Section 8.

\section{Modeling the Angular Power Spectra Of Light From External Galaxies}

The bandpowers in L10 are reported as ${\cal D}_\ell$, where
\be
{\cal D}_\ell \equiv \frac{\ell(\ell+1)}{2\pi} C_\ell.
\ee
In this work, we present constraints on measured powers in terms of either $C_\ell$ and ${\cal D}_\ell$ 
depending on the source.  
$C_\ell$ is constant for a Poisson distribution of sources, 
and is well suited to describing the Poisson point source power. 
On the other hand, ${\cal D}_\ell$ is nearly constant for the kSZ, tSZ, and clustered point 
source components at the angular scales of interest in this work.  
We generically refer to the tSZ, kSZ, and clustered contributions to the power spectra as 
``flat'' components due to their weak dependence of  ${\cal D}_\ell$ on $\ell$---weak compared to both the CMB temperature fluctuation power spectrum in the 
damping tail region and the Poisson power spectra. 

\subsection{Poisson}

The discrete nature of galaxies leads to fluctuations in their number in any comoving volume that are often approximated as following a Poisson distribution.  
Assuming this distribution, the resulting power spectrum depends on the flux distribution function as follows:
\be
\label{eqn:PoissonPower}
C_\ell = \int_0^{S_{\rm cut}} S^3 \frac{dN}{dS} d(lnS), 
\ee
and is independent of $\ell$.  
Here $(dN/dS) \Delta S$ is the number of sources per unit angular area in a flux bin of width $\Delta S$.  

\begin{figure*}
\begin{center}
\centerline{
\includegraphics[width=8.5cm]{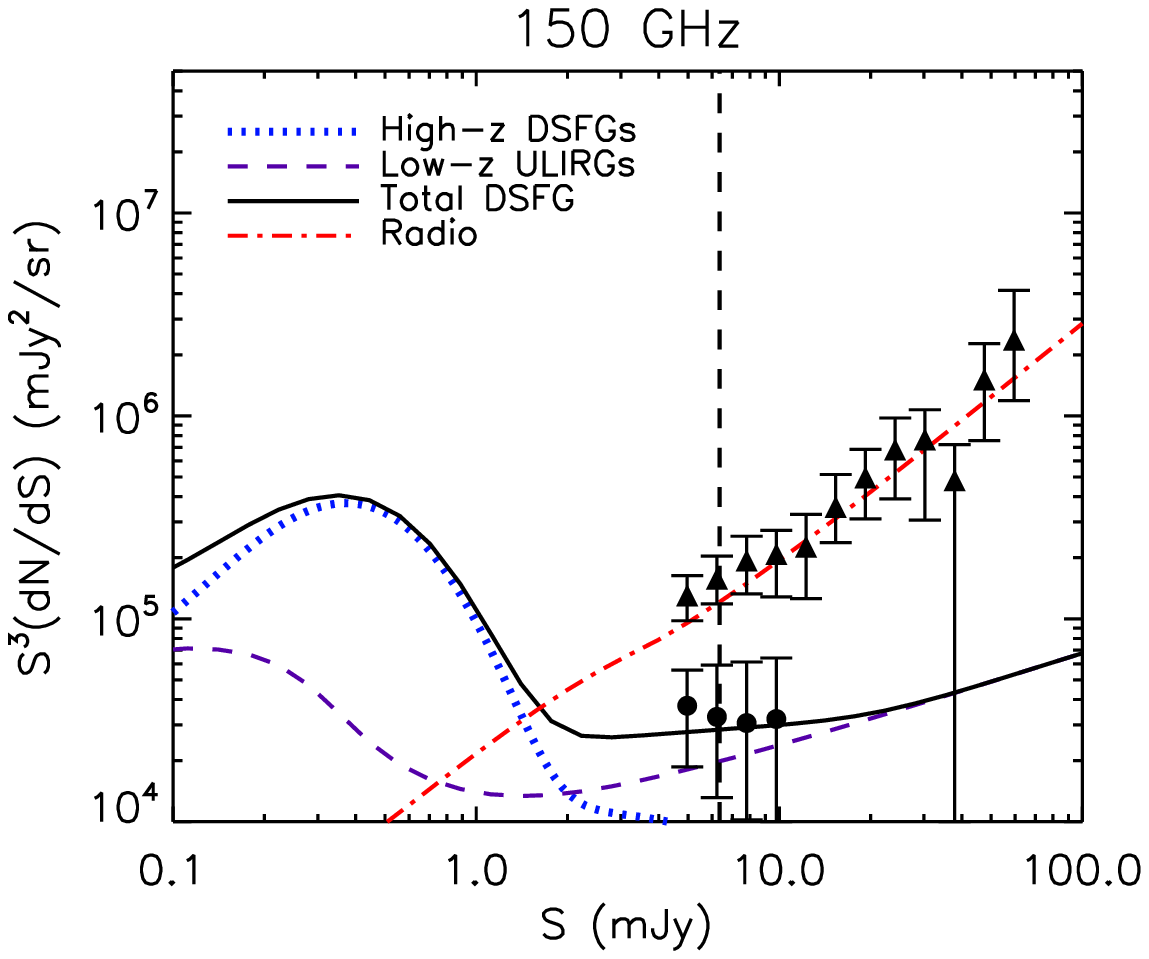}
\includegraphics[width=8.5cm]{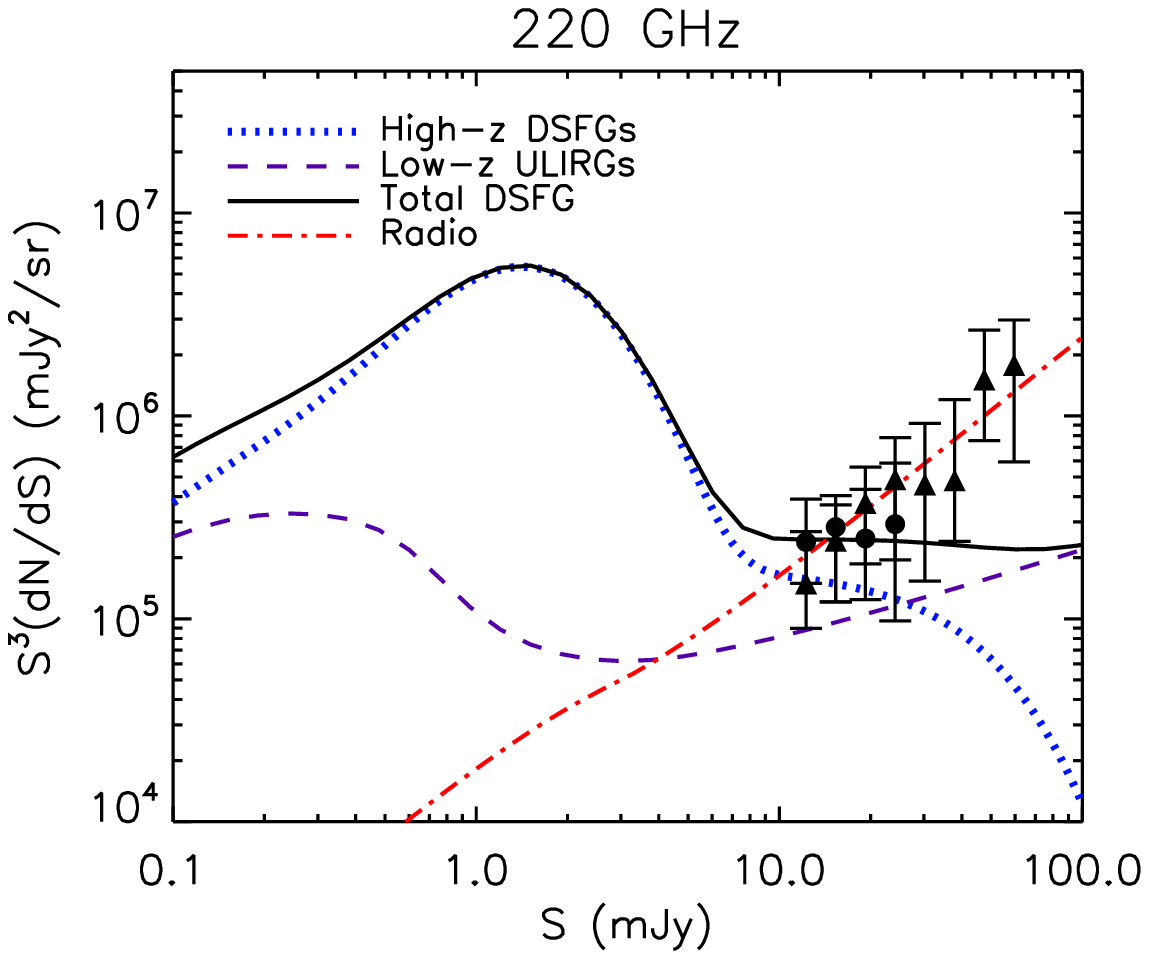}}
\end{center}
\caption{\small $S^3 dN/dS$ vs. flux ($S$) at $150\,$GHz ({\it left panel}) and $220\,$GHz ({\it right panel}).  
Source counts of \citet{negrello07} from high-$z$ ($z > 1$) DSFGs (\emph{blue dotted curve}), low-$z$ DSFGs (\emph{purple dashed curve}, often called ultra-luminous infrared galaxies (ULIRGs) due to their detection at IR wavelengths) and the sum of these two (\emph{black solid curve}).  The \citet{dezotti05} model radio galaxy curve is also plotted (\emph{red dot-dashed curve}), extrapolated from $150\,$GHz with $\alpha=-0.5$.  We extrapolate the Negrello model predictions as described in the text.  The \emph{vertical dashed line} indicates the flux threshold at $150\,$GHz set for masking of point sources from both maps.  At the bright end, the source counts of V09 due to radio galaxies (\emph{filled triangles}) and DSFGs (\emph{filled circles}) are also plotted.
\label{fig:s3dnds}}
\end{figure*}

In Figure~\ref{fig:s3dnds}, we plot the integrand in Equation~\ref{eqn:PoissonPower} for models and for the measured
source counts in V09.  
Note that with the logarithmic $x$-axis, the Poisson power is proportional to the area under the curve.  
By ``spectral index" in this work we mean $\alpha_{\nu_1-\nu_2}$ in the relation $S_{\nu_1} = S_{\nu_2} (\nu_1/\nu_2)^{\alpha_{\nu_1-\nu_2}}$.  
The DSFG source-count model predictions are from \citet{negrello07}\footnote{These source counts are based on the physical model of \citet{granato04} for high-$z$ SCUBA-like sources (that we call high-$z$ DSFGs in Figure~\ref{fig:s3dnds}) and on a more phenomenological  approach for late-type starburst plus normal spiral galaxies (that we call low-$z$ ULIRGs in Figure~\ref{fig:s3dnds}).} 
extrapolated to SPT frequencies with two broken power laws since the low-$z$ population is expected to have a spectral index that steepens with increasing frequency and the high-$z$ population is expected to have a spectral index that decreases with increasing frequency (see Figure~\ref{fig:DSFGseds}).  For the high-$z$ (low-$z$) populations, we use $\alpha_{220-350} = 3.2\,(3.0)$ and $\alpha_{150-220} = 3.5\,(2.0)$, with the shallower indices at low redshift due to
synchrotron and free-free emission.  The radio galaxy source counts of \citet{dezotti05} are also shown, extrapolated from 150 to $220\,$GHz with $\alpha_{150-220}=-0.5$.  The vertical dashed line in the left panel indicates the flux threshold used at $150\,$GHz to determine the point source mask that is used in both the 150 and $220\,$GHz SPT maps.  Also shown are source counts from SPT data (V09) separated into DSFG and AGN populations.  

According to the models, radio galaxies and DSFGs have very different source-count flux distributions.  While the radio galaxies dominate the SPT detections, their contribution to the Poisson power (in maps with detected sources  masked) is subdominant to those from the DSFGs.  Below $S \sim 3\,$mJy at $150\,$GHz, we see contributions from a large population of DSFGs extending in redshift from $z \sim 1$ to much higher redshifts.  We expand our consideration of radio galaxy contributions in Section~\ref{subsec:radiogal}.

\subsection{Clustering of dusty star-forming galaxies}
\label{subsec:clustering}

In this section, we calculate the clustering properties of DSFGs from models and compute the resulting power spectra.  
We consider two models in this work, the rich empirical model of \citet{lagache03}, as updated in \citet{lagache04}\footnote{In order to take into account constraints on number counts at 1.9, 4.3, and 12.5$\,$THz (160, 70, and 24$\,\mu$m) obtained from {\it Spitzer} observations \citep{papovich04,dole04}}, which we henceforth refer to as the LDP model, and a more simplistic ``single spectral energy distribution (SED)" model.  The advantage of the single-SED model is that it has a limited set of parameters which can be varied to match both the SPT and external data.
The template shapes that are used in the analysis of the SPT and BLAST band power data are presented
in Section~\ref{sec:analysis}.  
Examples of other efforts to model the power from clustered point sources include \citet[hereafter S09]{sehgal10} and \citet{righi08}.

Following \citet{haiman00} and \citet{knox01}, we calculate the clustering properties 
of the background light by assuming that the emissivity density, $j$, is
a biased tracer of the dark-matter density fluctuations, $\delta j/\bar j=b\delta \rho/\bar \rho$.
In a flat universe, the intensity, $I_\nu$, is related to the comoving emissivity density via
\be
I_\nu = \int dz \frac{d\chi}{dz} a \bar j(\nu,z)\left(1+\frac{\delta j(\nu,z)}{\bar j(\nu,z)}\right),
\ee
where $\chi$ is the comoving distance to redshift $z$ and $a=1/\left(1+z\right)$ is the scale factor.

For an angular power spectrum defined via
\be
\langle \delta I_{\nu,lm} \delta I_{\nu',l'm'} \rangle  =  C^{I}_{\nu\nu',l} \delta_{ll'}\delta_{mm'},
\ee
we get
\be
\label{eq:clustering}
 C^{I}_{\nu\nu',l} = \int \frac{dz}{\chi^2}\frac{d\chi}{dz} a^2 b^2 \bar j(\nu,z) \bar j(\nu',z) P_M(k,z)|_{k=l/\chi}.
\ee
Here, we have assumed a spatially constant and frequency-independent bias, $b$.  
We further assume that
\be
P_M(k) = P_0(k) G^2(z),
\ee
where $G(z)$ is the linear-theory density contrast growth factor normalized
to unity today and $P_0(k)$ is the power spectrum today.

Nonlinear effects (which we do not include) are expected to alter the shape and 
amplitude of the power spectrum on the relevant scales 
by boosting power on smaller angular scales.  
We have ignored these corrections due to our limited sensitivity to the shape of the clustered
power spectrum and its model dependence.    
However, the amplitude correction is important to keep in mind when interpreting our results.  
In the halo model of large-scale structure \citep{seljak00}, the power spectrum of galaxies is the sum of a one-halo term and a two-halo term, depending on whether it arises from a pair of galaxies residing in one halo or two different halos.  
The extra power at small angular scales, above expectations from linear perturbation theory, arises from the one-halo term.  
The bias factors of the halos in which the galaxies reside scale the amplitude of the two-halo term, and usually an effective bias is defined as a weighted average over the bias factors of all the contributing halos.  
The bias we define above is, therefore, different than a halo-model bias.  
For example, in the analysis of $860\,$GHz maps from BLAST, \citet{viero09} find $b=3.9 \pm 0.6$ for the \citet{lagache04} model using a linear-theory calculation, and an effective bias of $b=2.4 \pm 0.2$, using a halo model.  
Curiously, the shape of their power spectrum agrees better with the linear-theory shape than with any halo-model shape they manage to derive from their model.  

We express sky brightness fluctuations in terms of 
the departure from the mean CMB temperature that would give the equivalent 
change in brightness.  The power spectra are related via
\be
C_{\nu\nu',l} = 
\left(\frac{\partial B_\nu}{\partial T}\right)^{-1}\left(\frac{\partial B_{\nu'}}{\partial T}\right)^{-1} C_{\nu\nu',l}^I,
\ee
where $B_\nu$ is the Planck function, the derivatives of which are evaluated at $T = \bar T_{\rm CMB}$. 

The emissivity density is related to differential source counts via
\begin{eqnarray}
\bar j(\nu,z) & = & \left(1+z\right)\int_0^{S_{\rm cut}} dS S \frac{d^2N}{dSdz}  \\
   & = &  \left(1+z\right) \nonumber \\
   && \int dln(L)S(L,z)\Theta(S-S_{cut}) \frac{d^2N}{dln(L)\:dz} \left(\frac{d\chi}{dz}\right)^{-1},
\end{eqnarray}
where $\Theta(x-y)$ is the Heaviside function equal to 1 for $x < y$ and 0 for $x \ge y$.
We take $d^2N/(dln(L)\:dz)$ from the empirical LDP source-count model, designed to accommodate all observational constraints on source counts and Poisson fluctuation power from the mid-IR to submillimeter wavelengths as economically as possible.  In this model, there are two different spectral types: ``normal" and ``starburst" galaxies.  The latter undergo a rapid evolution of the luminosity function between $z=0$ and $z \sim 1$ and provide the large majority of the background light.  

\begin{figure}[ht]\centering
\includegraphics[scale=0.7]{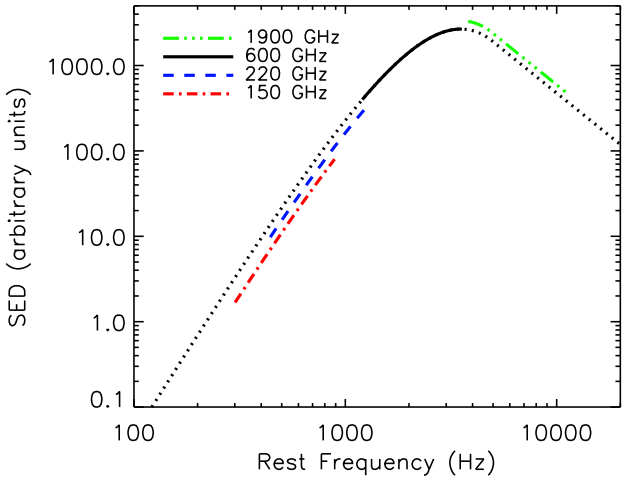}
 \caption[]{\small Assumed spectral shape of galaxies for our single-SED model of the
millimeter-wave background (\emph{black dotted}), with our fiducial parameter values $T_d=34$\,K, $\beta=2$, and $\alpha_{\rm mid-IR}=2$.  
Curves indicate the range of rest-frame frequencies of light emitted at redshifts between 1 and 5, for the observed frequencies of $1900\,$GHz (\emph{triple-dot-dashed green}), $600\,$GHz (\emph{solid black}), $220\,$GHz (\emph{dashed blue}), and $150\,$GHz (\emph{dot-dashed red}).}
\label{fig:sed1}
\end{figure}

In order to explore the broader implications of the SPT constraints on DSFG clustered power, we use an economical empirical model of the mean comoving emissivity density as a function of redshift.  
We use this ``single-SED model" rather than the LDP model since the LDP source counts are only publicly available for a fixed parameter set.  
In our single-SED model, we assume that all galaxies have the same greybody SED described by $f_{\nu}\propto\nu^{\beta}B_{\nu}\left(T_{d}\right)$ across the far-infrared to millimeter wavelengths. 
Here, $\beta$ is the emissivity spectral index of thermal graybody dust, $B_{\nu}$ is the Planck function, and $T_{d}$ is the effective dust temperature.  
We replace the mid-infrared exponential decline on the Wien side of the graybody with a power-law decline $f_{\nu}\propto\nu^{-\alpha_{\rm mid-IR}}$ by matching the two functions with a smooth gradient at the frequency $\nu^\prime$ that satisfies $d \ln f_{\nu^\prime}/d\ln\nu^\prime=-\alpha_{\rm mid-IR}$, as is done in \citet{blain03}.  
This replacement phenomenologically accounts for emission from dust with temperature greater than $T_d$, which becomes important beyond the peak of the emission from the dust with temperature $T_d$.  In Figure~\ref{fig:sed1}, we plot this SED for a specific choice of parameter values and show the range of rest frequencies probed by observations at SPT, BLAST, and {\it Spitzer} frequencies.  

\begin{figure*}
\begin{center}
\centerline{
\includegraphics[width=8.5cm]{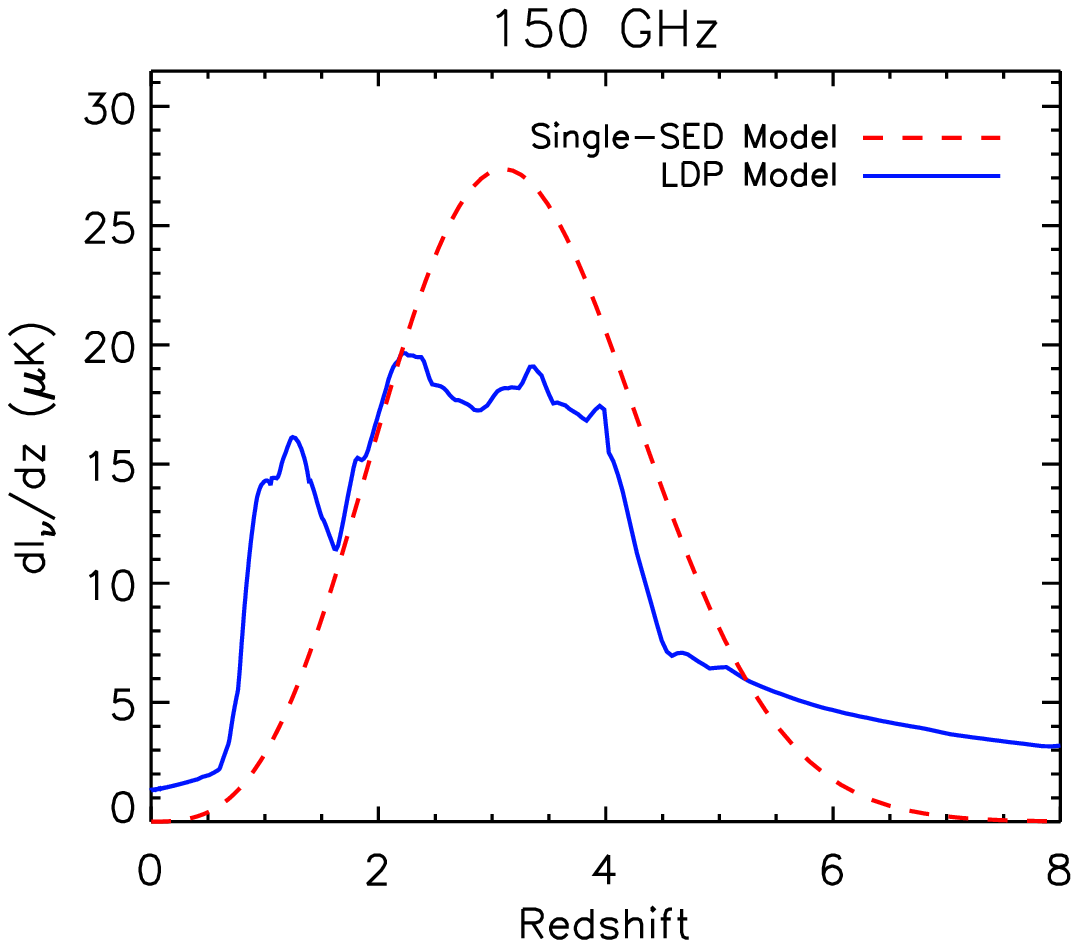}
\includegraphics[width=8.5cm]{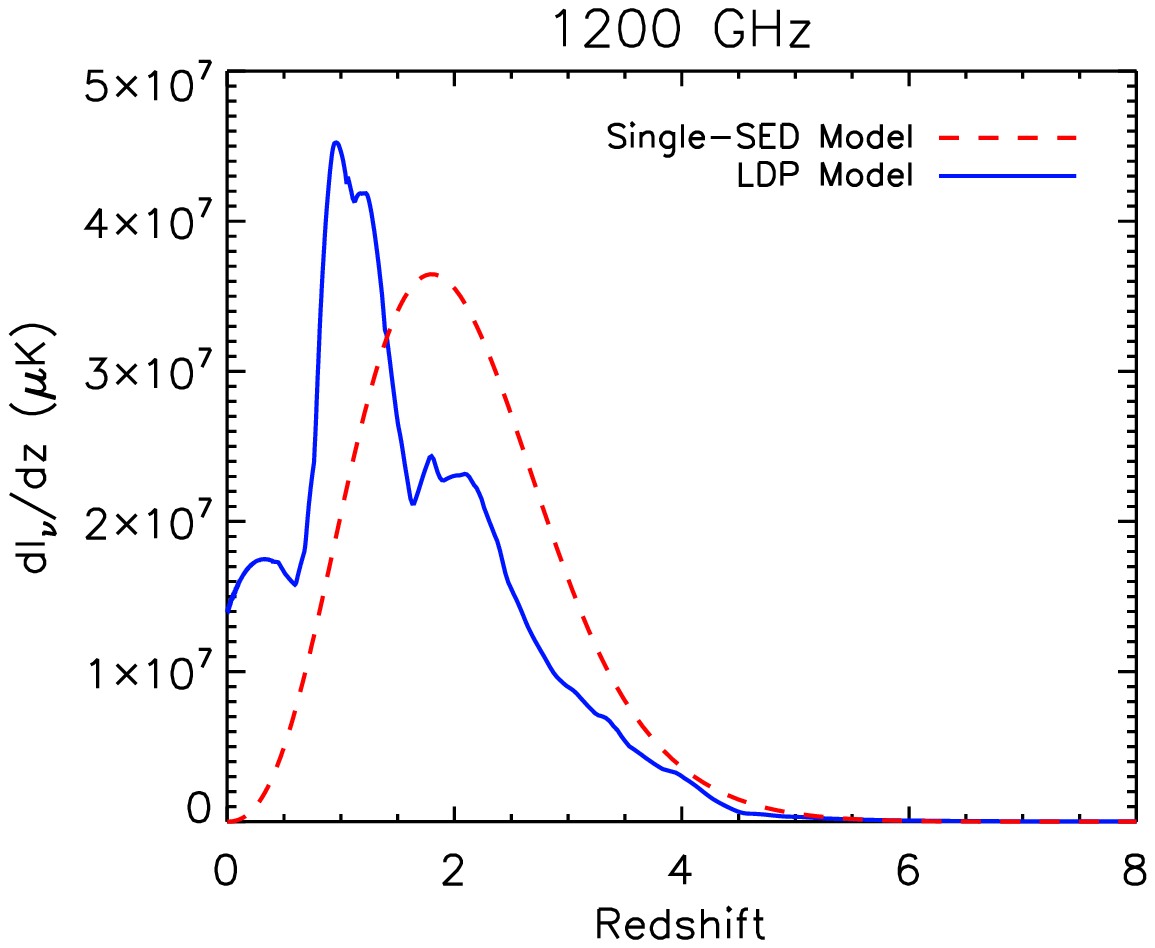}}
\end{center}
  \caption{\small Redshift-dependence of contributions to the mean background intensity, $d\bar I_\nu/dz$ at 150 and $1200\,$GHz ($\lambda = 250\, \mu$m) for the LDP model (\emph{blue solid curve}) and our fiducial single-SED model (\emph{red dashed curve}).  
  For these figures, the single-SED model is normalized, at both frequencies, to have the same $\bar I_\nu$ as the LDP model.
\label{fig:windows}}
\end{figure*}

Rather than parameterize a luminosity function, we directly parameterize the mean comoving emissivity 
density since that is all we need to make predictions of the clustered power.  
We write it as
\be
\bar j(\nu,z) \propto a \chi^2(z) \exp\left(-\frac{(z - z_c)^2}{2\sigma_z^2}\right) f_{\nu\left(1+z\right)}.
\ee
In Figure~\ref{fig:windows}, we show the contribution to the mean intensity
as a function of redshift, $d\bar I_\nu/dz = a \bar j(\nu,z) d\chi/dz$, 
at 150 and $1200\,$GHz ($250\, \mu$m) for both the LDP model and our fiducial single-SED model.   
 We choose $T_d=34\,$K, $z_c=2$, $\sigma_z=2$, $\beta=2$, and $\alpha_{\rm mid-IR}=2$ as our fiducial single-SED model parameters for reasons given in Section~\ref{sec:DSFGmodels}.  
 From Figure~\ref{fig:windows}, we see that the redshift at which
$d\bar I_\nu/dz$ peaks decreases with increasing observing frequency, so the power seen by BLAST is predominantly coming from lower redshift than 
that seen by SPT.

\begin{figure}[ht]\centering
\includegraphics[scale=0.7]{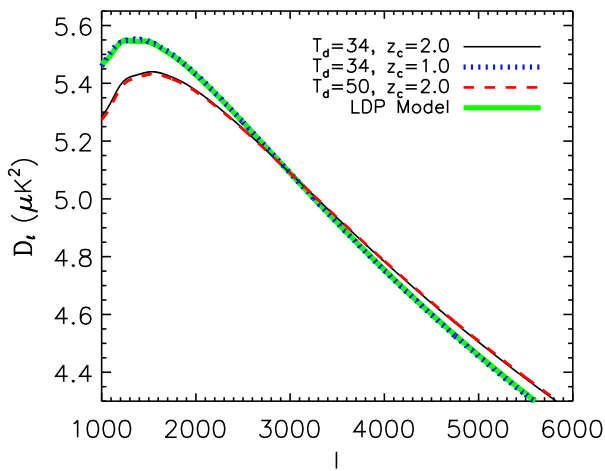}
  \caption[]{\small Dependence on model parameters of the shapes of the angular power spectra of clustered point sources for our single-SED model at $150\,$GHz for various parameter combinations.  
  We show three curves all with $\sigma_z=2$, $\beta=2$, and $\alpha_{\rm mid-IR}=2$ and with the following $T_d$ and $z_c$ combinations: $T_d=34$\,K and $z_c=2$ (\emph{thin, solid, black line}), $T_d=34$\,K and $z_c=1$ (\emph{dotted, blue line}), $T_d=50$\,K and $z_c=2$ (\emph{dashed, red line}).  
  Also shown for comparison is the shape of the LDP model (\emph{thick, solid, green line}).  
  All power spectra are normalized to the $T_d=34$\,K and $z_c=2$ model at $\ell = 3000$.  }
\label{fig:compparams}
\end{figure}

In Figure~\ref{fig:compparams}, we show how power spectrum shapes at $150\,$GHz change with adjustments to $T_d$ and $z_c$.  
We also show how they change with a switch from the single-SED model to the LDP model.  All four curves are normalized to the same value at $\ell=3000$. 
The relative percent difference in power between single-SED model curves for various parameter values is small compared to the accuracy with which we can measure the power.  
 Thus, we fix the template shapes at each frequency, calculated with our fiducial single-SED model, and allow only the amplitude to vary in this analysis.

\subsection{Clustered Radio Galaxies}
\label{subsec:radiogal}

Models of galaxy evolution for both radio galaxies and DSFGs generically predict a clustering amplitude on the order of a few percent of the intensity of the mean background on arcminute scales. 
Therefore, the amplitude of the mean background is the quantity of importance.  
The mean temperature due to undetected radio galaxies is difficult to estimate, and will be model dependent.  
However, there are strong constraints from lower-frequency experiments.  
Detailed models by \citet{gervasi08} predict that the mean brightness temperature due to faint radio sources is $230\, {\rm mK} (\nu/{\rm GHz})^{-2.7}$ at low frequencies, while recent results from the 
ARCADE2 experiment \citep{fixsen09} find the mean extragalactic temperature above the CMB 
at frequencies from $3$ to $10\,$GHz to be $(1260\pm 90)\, {\rm mK} \,(\nu/{\rm GHz})^{-2.6}$.  
The discrepancy between source models and the observed background is currently unresolved \citep{seiffert09}, but as an upper limit on the possible contamination from radio galaxies we can extrapolate the observed ARCADE2 temperature from their observed range of $3-10\,$GHz to $150\,$GHz, finding $T^{RJ}_{\rm radio} \sim 2.7 \,\mu {\rm K}$ in antenna temperature, or about 5$\,\mu {\rm K}$ in CMB temperature units.
 Assuming that 10\% of this is clustered, the power contribution from clustered radio sources is about 0.25~$\mu {\rm K}^2$.

Based on small-scale CMB anisotropy measurements at $30\,$GHz from the Sunyaev--Zel'dovich Array (SZA) \citep{sharp10} and the Cosmic Background Imager (CBI) \citep{sievers09}, the above value of $0.25\,\mu {\rm K}^2$ is almost certainly an overestimate of the contribution from clustered radio sources. 
In radio source models, the dominant contribution to the mean extragalactic background comes from faint sources, although there are substantial contributions from a broad range in flux.  
Extrapolating the ARCADE2 results to $30\,$GHz predicts the power from clustered radio sources to be of order 300$\,\mu {\rm K}^2$. This prediction is well above the SZA $95\%$ confidence upper limit of 149 $\mu {\rm K}^2$ and would account for the entire measured CBI excess power.  
Assuming that the SZA upper limit is entirely due to clustered radio galaxies yields a clustered radio contribution of roughly 0.1 $\mu {\rm K}^2$ at $150\,$GHz.  
Extrapolations from measured source counts predict a more reasonable contribution of $\sim 6\,\mu {\rm K}^2$ from clustered radio galaxies at $30\,$GHz (i.e., a factor of 50 lower than that inferred from the ARCADE2 extrapolation), which translates into much less than 0.01~$\mu {\rm K}^2$ in power at $150\,$GHz.  
We can, therefore, safely neglect the clustering of radio sources.

\subsection{Unclustered Radio Galaxies}
\label{subsec:unclustradio}

Unlike DSFGs, the power contributed by radio sources is expected to be dominated by brighter sources; 
the number counts as a function of flux are such that the quantity $d\bar I_\nu/d\ln{S} =S^2 dN/dS$ 
increases toward higher flux \citep{dezotti05}.  
The detected radio sources in the SPT maps are consistent with the expected distribution as shown in V09 
and here in Figure~\ref{fig:s3dnds}.  
Applying Equation~\ref{eqn:PoissonPower} to the \citet{dezotti05} source counts, we calculate the expected 
power from radio sources below the $150\,$GHz SPT $5\sigma$ flux threshold of $6.4\,$mJy, to be 
$C_l=7.9 \times 10^{-7}\,\mu {\rm K}^2$.  We also examine the power spectrum of the all-sky radio 
source maps produced by S09 and find a very similar levels of residual radio source power 
($C_l=8.0 \times 10^{-7}\,\mu {\rm K}^2$) after masking sources above $6.4\,$mJy.  
This is about 11\% of the power we expect from DSFGs and is a significant source of uncertainty 
in our measurement of the mean spectral index of the DSFGs which we consider in Section~\ref{sec:DSFGmodels}.

\subsection{Correlations with Galaxy Clusters}
\label{subsec:psclustercorrelations}

A possible concern for results derived from the $150\,$GHz autospectrum is that radio galaxies and DSFGs are correlated with galaxy clusters. 
In this case, the SZ effect from the galaxy clusters can ``fill in'' flux from the sources and 
lead to a suppressed signal. 
There are several lines of reasoning that argue against significant radio correlations; for example, the works by \citet{lin09} and S09 show the number of cluster-correlated radio sources is expected to be small.  
We also test this assumption using the simulated tSZ and radio source maps produced by S09 for $148\,$GHz. 
We compute the power spectra at $\ell = 3000$ for the tSZ map, the tSZ+radio source map, and the radio source only map after masking sources above $6.4\,$mJy.  We find an anti-correlation coefficient of 2.3\% for  the two components which agrees well with other lines of argument. 

We also do not expect significant correlations between galaxy clusters and DSFGs.
Galaxy clusters are not observed to be strongly overdense in DSFGs relative to the field (a factor of 20, while the mass is overdense by a factor of 200 \citep{bai07}), so only a small fraction of the galaxies contributing to the millimeter background can live in galaxy clusters\footnote{Note that this latter argument is inconsistent with the modeling of DSFGs in S09.  In their halo model, massive galaxy clusters would be expected to have roughly 200 times the density of sources in the field, since the number of galaxies roughly tracks the mass of the halo.}.  

\section{Modeling the Other Signals}

The clustered power and the Poisson power are not the only significant contributors at 150 and $220\,$GHz 
for the angular scales of interest.  In order to interpret the data, we also need to model fluctuation power from primary CMB anisotropy, the tSZ effect, the kSZ effect, and Galactic cirrus.

\subsection{CMB}

We use {\textsc CAMB} \citep{lewis00} to predict the primary CMB anisotropy for the standard, six-parameter, spatially flat, lensed $\Lambda$CDM cosmological model. 
The six parameters are the baryon density $\Omega_b$, the density of cold dark matter $\Omega_c$, the optical depth to recombination $\tau$, the angular size of the sound horizon on the surface of last scattering $\Theta$, the amplitude of the primordial density fluctuations $ln[10^{10} A_s]$, and the scalar spectral index $n_s$.  
We have not explored alternative cosmological models, although such models could affect the point source results presented in this work by modifying the predicted high-$\ell$ primary CMB anisotropy.
 
Gravitational lensing of primary CMB anisotropies by large-scale structure tends to increase the power at small angular scales which could impact the measurement of the clustered and Poisson point source contributions. 
Including lensing out to $\ell=10,000$ proved computationally challenging.  We avoid this computational constraint by calculating the lensing contribution for the best-fit CMB cosmological model, and adding this estimated lensing contribution to the unlensed CMB power spectra calculated for each parameter set.  Given the small range of $\Omega_m$ allowed by the combination of WMAP5, ACBAR, and QUaD data, we determine that using a fixed lensing contribution will misestimate the actual lensing by no more than 30\%.  The lensing contribution to the high-$\ell$ CMB spectrum is $\sim$1.5$\,\mu {\rm K}^2$ near $\ell = 3000$, and falls at higher $\ell$.  This 0.5$\,\mu {\rm K}^2$ lensing error ($30\%$ of $1.5\,\mu {\rm K}^2$) is smaller than the clustered power we detect at $150\times220$ and $220\times220$ by factors of 40 and 140 respectively.
 
We also use {\textsc EmuCMB}\footnote{Software and results of validation testing are available at {\textsc http://www.emucmb.info}.} to calculate the lensed primary CMB anisotropy.  
EmuCMB is a CMB Boltzmann code emulator that can return a lensed CMB temperature power spectrum out to $l=5000$ thousands of times faster than {\textsc CAMB}.  
We use it for the subtraction of the CMB contribution to the bandpowers in order to
study residuals of our model fits, as described below.  

\subsection{Thermal SZ}

The tSZ template we use is based on the simulated tSZ maps released by S09 for a WMAP5 cosmology. 
There is significant theoretical uncertainty in the shape and amplitude of the tSZ effect for a given cosmology. 
The amplitude uncertainty is mitigated by making the normalization of the tSZ template a free parameter at $150\,$GHz.  We assume that the tSZ contribution to the $150\times220$ and $220\times220$ bandpowers is zero which should be an excellent approximation at current significance levels.

\subsection{Kinetic SZ}

Galaxy clusters and other large-scale structures carry line-of-sight velocities that Doppler shift scattered CMB 
photons, resulting in a CMB temperature anisotropy known as the kinetic SZ effect \citep{sz1980}.  
The signal is proportional to the product of the free electron density and velocity and has a spectrum identical to a variation of the primary CMB blackbody temperature.  
In contrast to the tSZ effect, electrons with temperatures as low as 10$^4\,$K contribute to the kSZ effect. 
Therefore higher redshifts, before massive objects finish collapsing, have a larger contribution to the kSZ effect.
Clusters of galaxies at low redshift dominate the kSZ power on small angular scales, while high-redshift reionizing regions that are several tens of Mpc across have their largest relative contribution on angular scales around $\ell \sim 2000$, close to where secondary
anisotropies and point sources begin to dominate the anisotropy power.
This so-called patchy reionization contribution \citep{gruzinov98,knox98} is highly dependent on details of reionization and hence is  highly uncertain.  

The kSZ contribution is an important source of uncertainty in our determination of power from clustered DSFGs.  
We first analyze the SPT bandpowers with a baseline assumption for the kSZ power, 
and then discuss the impact of uncertainty in the assumed kSZ signal on our conclusions.  
The baseline model is taken from the treatment of S09, and includes post-reionization contributions to kSZ calculated for WMAP5 best-fit cosmological parameters.
Patchy reionization may contribute significantly to the kSZ power spectrum and, therefore, we consider the implications of this potential additional kSZ power on our conclusions.  

\subsection{Galactic Cirrus}
\label{sec:cirrus}

Another source of fluctuation power is emission from the cirrus-like dust clouds in our Galaxy.  Although the area of sky used in this analysis has a relatively low level of emission from Galactic cirrus as measured by the {\it Infrared Astronomical Satellite} ({\it IRAS}) \citep{neugebauer84}, this emission is present at some level and its effect on the SPT spectra must be quantified.  To this end, we cross-correlate the $150\,$GHz and $220\,$GHz SPT maps with predictions for cirrus emission at these frequencies from model eight of \citet[hereafter FDS]{finkbeiner99}.  We detect correlated power at high significance at both frequencies.  The correlated power displays a $C_\ell \propto \ell^{-3.5}$ shape after being corrected for the transfer functions of SPT and FDS.  We use the normalization of the correlated power to predict the level of cirrus power in the SPT spectra.  In practice this is implemented as a prior on the normalization of a $C_\ell \propto \ell^{-3.5}$ term in the multi-component fit to the SPT power spectra.  We find that the inclusion of this term has a negligible effect on all components except for the amplitude of the clustered point source power, which decreases by 8\% (or 0.3$\sigma$) at $150\times220$ and 7\% (or 0.2$\sigma$) at $220\times220$.

We note that the FDS predictions are based on $100\,\mu$m emission and extrapolated to millimeter wavelengths using dust temperature estimates from the Diffuse Infrared Background Experiment (DIRBE) on the {\it COBE} satellite.  DIRBE  has degree-scale resolution, so if there is significant structure in the dust temperature at smaller scales, there will be some loss of correlation between the FDS template and the true cirrus emission in the SPT bands. This could bias our prediction for cirrus power at 150 and $220\,$GHz low.  In order to estimate the magnitude of this effect, we have studied the angular power spectrum of the dust temperature in the FDS model and extrapolated it to smaller angular scales.  
We find that temperature structure could increase the predicted cirrus power by up to 10\%; such a small change in our prior on the cirrus power would have a negligible effect on our results, so we choose to ignore any decorrelation effects from temperature structure.

\section{Data Collection and Determination of Power Spectra} 

We use data from one $100\,$deg$^2$ field centered at right ascension $5^\mathrm{h} 30^\mathrm{m}$, declination $-55^\circ$ (J2000) observed by SPT in the first half of the 2008 austral winter.  
We use the central $78\,$deg$^2$ of the map which has uniform noise at a depth of $18\,\mu {\rm K-arcmin}$ at $150\,$GHz and $40\,\mu {\rm K-arcmin}$ at $220\,$GHz. Point sources detected at $> 5\sigma$ ($6.4\,$mJy) at $150\,$GHz are masked from both maps in the analysis.

From this two-frequency map, L10 estimate bandpowers for the $150\,$GHz auto spectrum, the $220\,$GHz auto spectrum, and the $150\times220\,$GHz cross-spectrum.  
The bandpowers are estimated with a pseudo-$C_\ell$ method. 
A cross-spectrum based analysis is used to eliminate noise bias \citep{polenta05, tristram05}. 
Corrections for the time-stream filtering, beams and finite sky coverage are applied using the framework of the MASTER formalism \citep{hivon02}.  
Bandpower uncertainties are determined using signal-only simulations to estimate sample variance, and using the scatter between disjoint sets of cross-spectra to estimate the instrumental noise variance.
A full description of the bandpower estimation may be found in L10 and the bandpowers and window functions may be downloaded from the SPT Web site.\footnote{http://pole.uchicago.edu/public/data/lueker09/}

\section{Analysis of the three power spectra}
\label{sec:analysis}

\begin{figure*}
\begin{center}
\centerline{
\includegraphics[width=17cm]{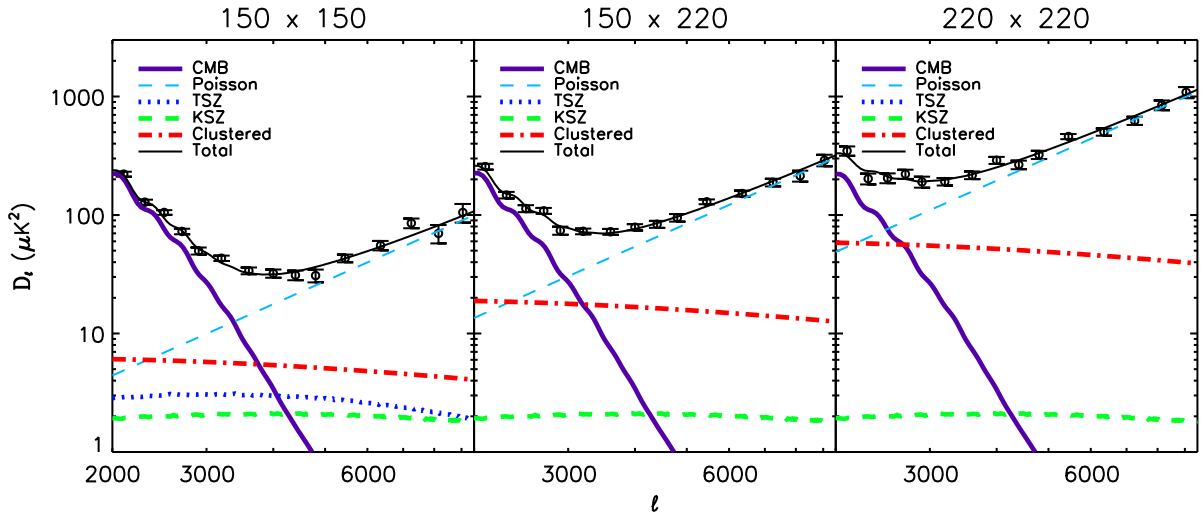}}
\end{center}
\caption{\small The SPT bandpowers (\emph{black circles}) are plotted at each frequency along with the following model components:  total power (\emph{thin, solid, black line}), CMB (\emph{thick, solid, purple line}), tSZ effect(\emph{dotted, blue line}), clustered point source power (\emph{dot-dashed, red line}), Poisson point source power (\emph{thin, dashed, cyan line}), kSZ effect (\emph{thick, dashed, green line}), and Galactic cirrus (\emph{thick, tripple-dot-dashed, orange line}).  
In all cases, the kSZ effect is fixed to the S09 model.
At  $150\times220$ and $220\times220$, the plotted models are the maximum-likelihood solutions.
We also plot the maximum-likelihood CMB and Poisson point source components for the $150\times150$ bandpowers; however, the other components plotted at $150\times150$ are extrapolated from other data.
The tSZ component is set to the L10 maximum-likelihood amplitude for this kSZ model.  
The clustered point source component at $150\times150\,$GHz is fixed to the value inferred by extrapolating from the best-fit amplitudes of the other two power spectra assuming perfect correlation.
The cirrus component at $150\times150\,$GHz is fixed to the value predicted by cross-correlating with FDS model eight.
\label{fig:powspec_model}}
\end{figure*}

\begin{figure*}
\begin{center}
\centerline{
\includegraphics[width=5.8cm]{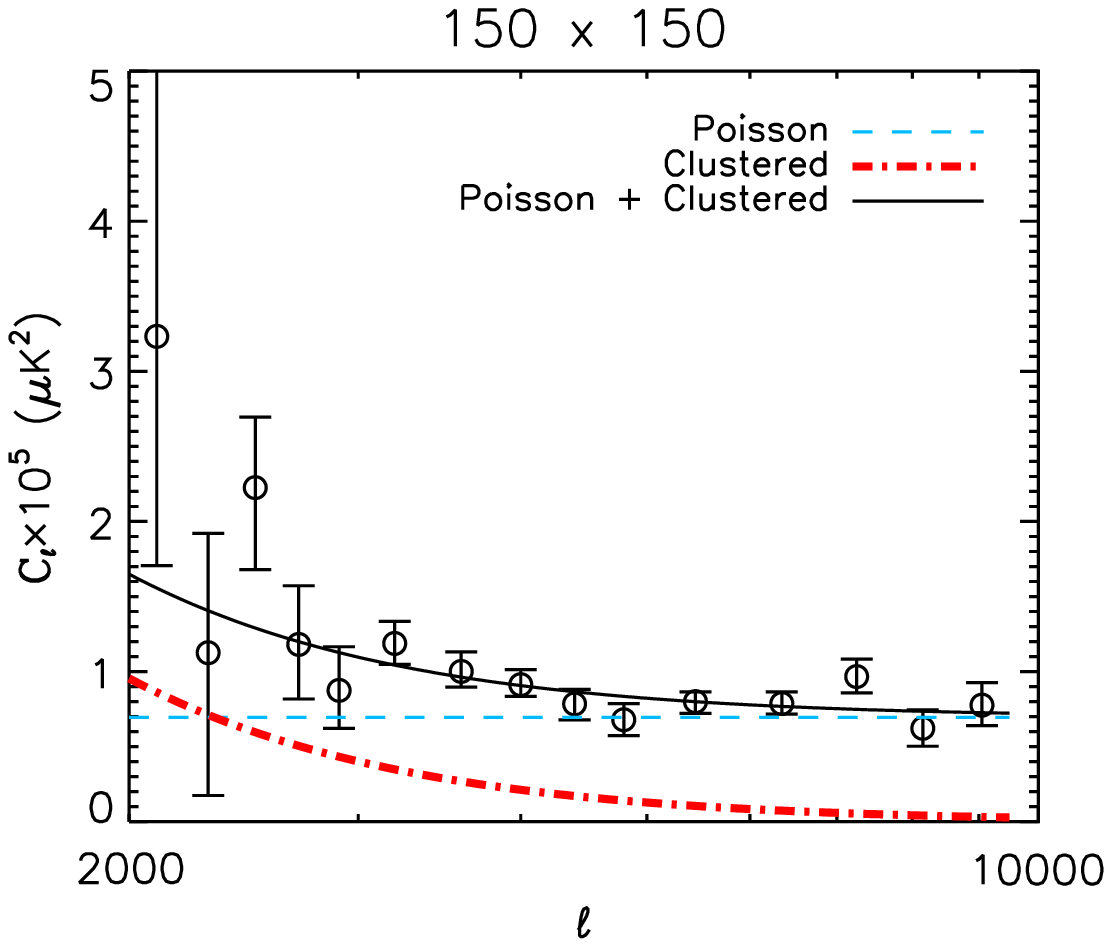}
\includegraphics[width=5.8cm]{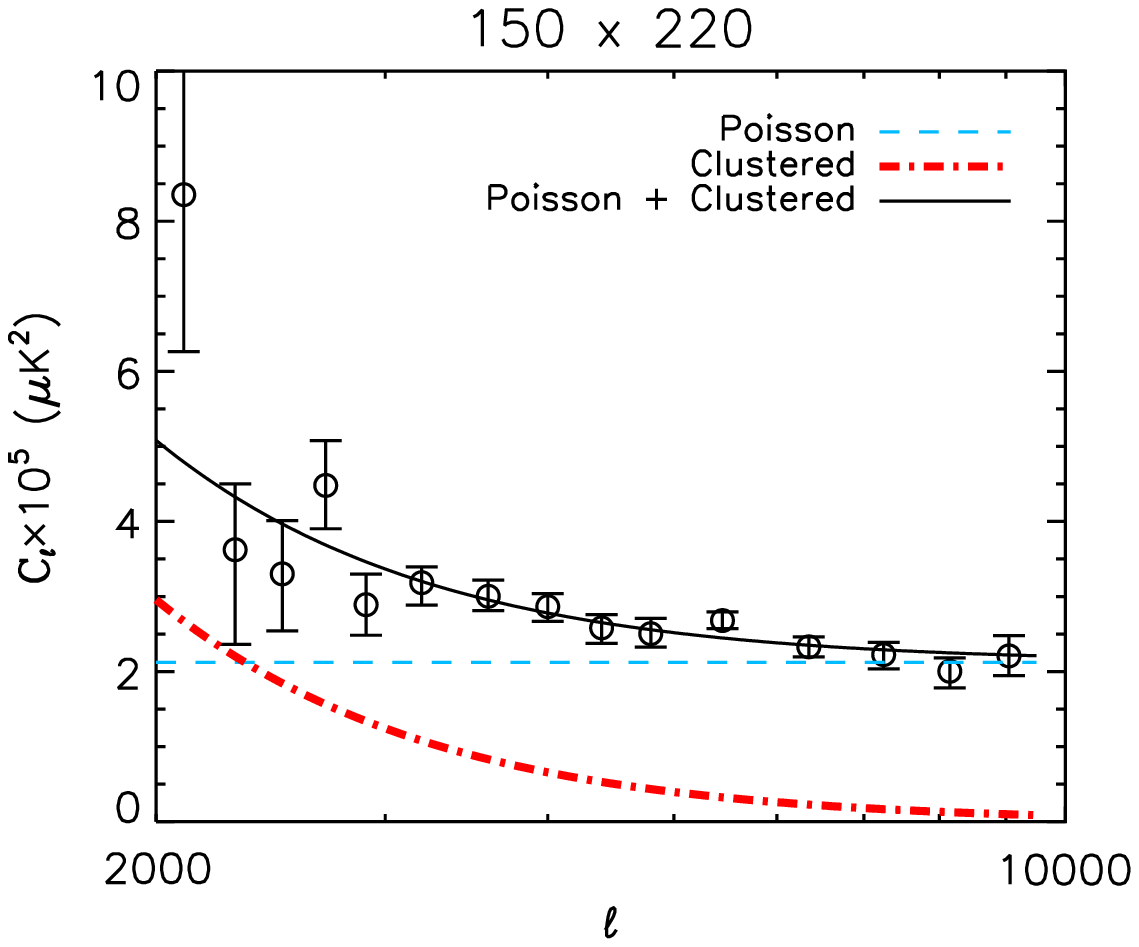}
\includegraphics[width=5.8cm]{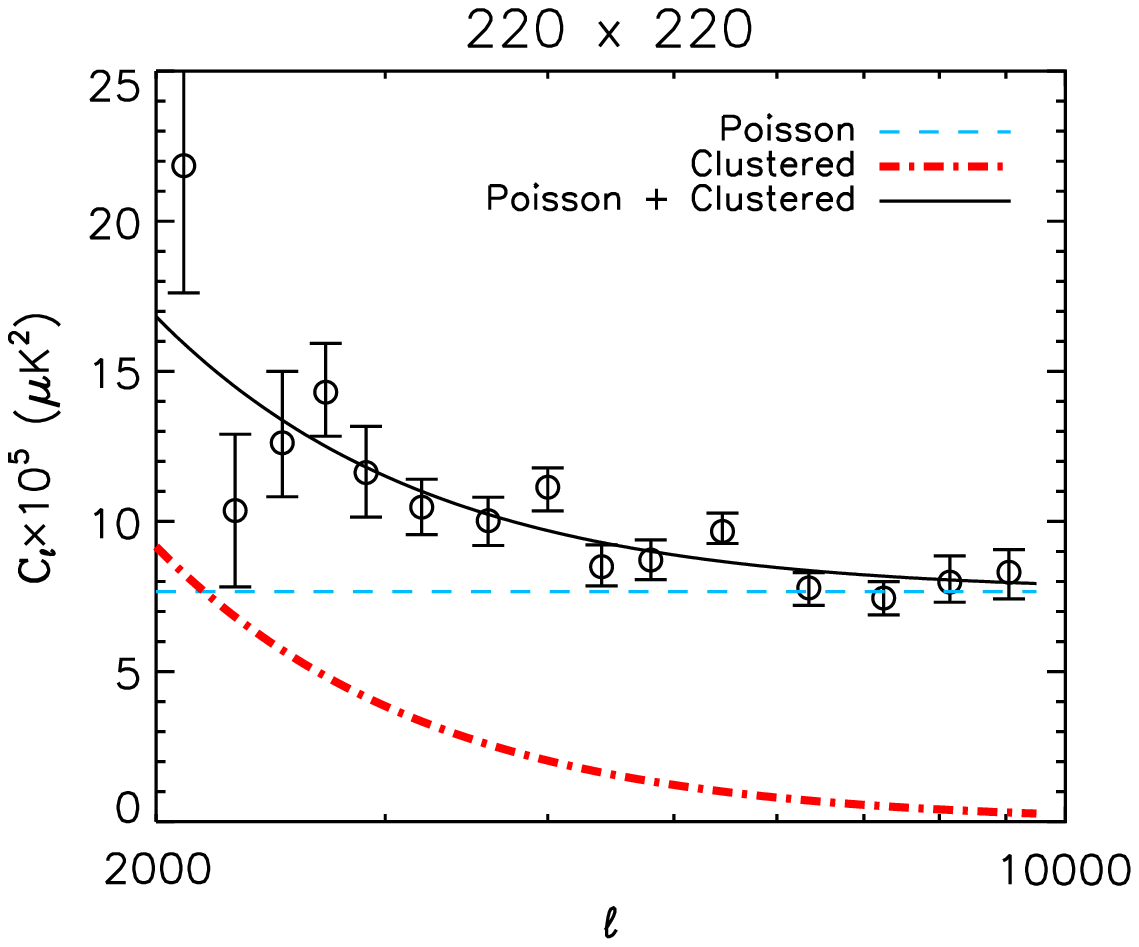}}
\end{center}
\caption{\small The residual SPT bandpowers (\emph{black circles}) after subtraction of the tSZ, kSZ, CMB, and cirrus model components shown in Figure~\ref{fig:powspec_model}.  Note that the power is plotted in $C_\ell \propto D_\ell/(\ell(\ell+1))$ as the Poisson point source power is flat in $C_\ell$.
The residual model components are as follows (see the description in the Figure~\ref{fig:powspec_model} caption): total residual power (\emph{thin, solid, black line}), clustered point source power (\emph{thick, dot-dashed, red line}), and Poisson point source power (\emph{thin, dashed, cyan line}).  
The error bars have been expanded to include uncertainties in the CMB contribution, as described in the text.  
\label{fig:powspec_resid}}
\end{figure*}

We show the bandpowers for the three power spectra and maximum-likelihood models for the $150\times220$, and $220\times220$ power spectra in Figure~\ref{fig:powspec_model}.  For $150\times150$ the tSZ, kSZ, and clustered point source power spectra are ``best-guess'' models as explained below.  The model includes the primary CMB, tSZ, kSZ, cirrus, and both Poisson and clustered DSFG components.  In order to better visualize the point source contribution and need for a clustered component, we also plot the bandpowers after subtracting the estimated primary CMB, SZ, and cirrus contributions in Figure~\ref{fig:powspec_resid}.  The bandpowers in Figs.~\ref{fig:powspec_model} and \ref{fig:powspec_resid} have been corrected for an $\ell$-independent calibration factor and an $\ell$-dependent beam factor (see L10) for comparison with the models.  Calibration and beam uncertainties have not been included in the error bars.

Strictly for visual inspection (to guide the ``$\chi$ by eye''), the residual bandpower error bars 
in Figure~\ref{fig:powspec_resid} have been expanded to contain the uncertainty in the subtracted CMB contributions. This uncertainty is estimated by examining the scatter in a random sampling of CMB models from the parameter chains.  Fixed tSZ (only at $150 \times 150$), kSZ, and cirrus templates are also subtracted.
This method is approximate in that we have ignored possible correlations between these various components in doing the subtraction.

From the center and right panels of Figure~\ref{fig:powspec_model}, it is clear that we need the clustered component (or something like it) to avoid having an excess of power at low $\ell$ for the $150\times220$ and $220\times220$ bandpowers.   
For a single spectrum, the low-$\ell$ excess could be explained by a sufficiently large kSZ effect. 
However, the relative powers seen in the $150\times150$, $220\times220$, and $150\times$220 bandpowers are well fit by a source with a dusty rather than a kSZ frequency dependence.

Ideally, we would perform a simultaneous analysis of all three power spectra, which would help in discrimination between the numerous components with similar spatial properties.  It is clear from Figure~\ref{fig:powspec_model} that such an analysis would give strong quantitative constraints.  A multifrequency analysis with its additional challenges is planned for future work.  In this work, we explore constraints that we can set with the individual band powers.   First, we consider constraints from the $150\times150$ bandpowers alone.

The expected contribution of tSZ and kSZ to our $150\times150$ bandpowers complicates our analysis and ultimately prevents a definitive detection of clustered point source power at $150\times150$.  The similarity of the assumed $\ell$-space shapes of these three components leads to a near degeneracy in their amplitudes when constrained by the $150\times150$ bandpower data alone.  Consequently, we perform an analysis with just the clustered point source power spectrum template, and treat it as a proxy for the power from all three contributions combined.  We can interpret the constraints on the sum of these powers as an upper limit for any one of the signals. 
We find for the clustered power at $\ell=3000$
\be
\frac{\ell(\ell+1)}{2 \pi}C_{150\times150,l}^{\rm PS,C} < 14 \ \mu {\rm K}^2
\ee
at 95\% confidence.  There are slight differences in the template shapes that 
lead to source-dependent upper limits.  Using the S09 kSZ power spectrum as
a template, we find the power at $\ell=3000$ to be $<13\,\mu {\rm K}^2$.\footnote{Our upper limits for both the kSZ shape and clustered point source power shapes are the same if specified at $\ell = 3800$.}   
The analysis of L10 uses a DSFG-subtracted map and the assumption that the kSZ signal is given by S09 to produce the tighter 
upper limit for the kSZ power at $\ell=3000$ of
\be
\frac{\ell(\ell+1)}{2 \pi} C_{150\times150,l}^{\rm tSZ}< 6.9\ \mu {\rm K}^2
\ee
at 95\% confidence.  These estimates and upper limits are tabulated in Table~\ref{tab:resultsflat}.  For these upper limits, we conservatively ignore the contribution from cirrus.

In order to produce the upper limit on the clustered point source power at $150\times150$, we effectively assumed that the point source fluctuations are {\em not} highly anti-correlated with total SZ.
In the case of a large anti-correlation, it would be possible for the addition of sources to decrease the total fluctuation power.  
We might expect some degree of anti-correlation since more massive clusters will have increased (negative) tSZ flux and more member galaxies, some fraction of which will have dust and radio emission.  
As described in Section~\ref{subsec:psclustercorrelations}, the simulations presented by S09 have been used to constrain the correlation of tSZ with residual radio emission to a 
negligible level.  L10 also argue that the anti-correlation of tSZ with dust emission is sufficiently small to have negligible impact on our upper limits.  

In addition to constraining the total power, we also make a ``best-guess'' at the amplitude of each of these three signals in the $150\times150$ data.  
We assume the kSZ model from S09.  
For the tSZ effect, we take the best-fit tSZ amplitude for this kSZ model from L10.
The best-guess clustered point source power is an extrapolation from the best-fit clustered point source power values at $220\times220$ and $150\times220$ assuming the S09 kSZ model and a correlation coefficient of unity between the spatial fluctuations due to clustering at 150 and $220\,$GHz. 
We assume the predicted level of cirrus discussed in Section~\ref{sec:cirrus}.
These estimates produce the tSZ, kSZ, clustered point source, and cirrus curves plotted in the left panel of Figure~\ref{fig:powspec_model}.  
We subtract these tSZ, kSZ, and cirrus models as well as the best-fit CMB model from the $150\times150$ bandpowers to produce the left panel of Figure~\ref{fig:powspec_resid}.  

\begin{table*}[ht!]
 \begin{center}
 \caption{\label{tab:resultsflat} Summary of SZ Results}
 \small
 \begin{tabular}{c|c}
 \hline\hline
\rule[-2mm]{0mm}{6mm}
 Signal & Measured ${\cal D}_{3000}$ ($\mu {\rm K}^2$)\\
 \hline
 \hline
Upper limit on kSZ  &$<$ 13 (95\%)  \\
\hline
\hline
 DSFG-subtracted & \\
 tSZ + 0.46 kSZ & 4.2 ($\pm$  1.5) \\
\hline
 tSZ & 3.3 ($\pm$ 1.5) \\
\hline
Residual clustering &  $ < 0.3$ (95\%) \\
\hline
\hline
Fiducial kSZ & 2.0  \\
\hline\hline
\end{tabular}
\tablecomments{Upper limits and best-fit values evaluated at $\ell = 3000$ for tSZ, kSZ, and clustered power at $150\,$GHz.  
The model templates for all three components are essentially completely degenerate at $150\,$GHz. 
The first row has the flat power determined from the $150\,$GHz bandpowers which consists of the sum of all three components. 
We report the upper limit on kSZ power at 95\% confidence from the $150\,$GHz maps in the second row, derived under the assumption that all flat power is kSZ.
The next three rows are results from the DSFG-subtracted maps in L10 which are listed here for comparison. 
The subtraction technique reduces the kSZ relative to the tSZ by a factor of $\sim 2$ and nearly eliminates clustered DSFG power as discussed in Section~\ref{sec:DSFGjust}. 
The implied tSZ level under the assumption of the fiducial kSZ model is found on the fourth row. 
The upper limit at 95\% confidence on residual clustered power in the DSFG-subtracted maps is on the fifth row. 
The fiducial kSZ model power is shown in the last row for reference.
We conservatively ignore the expected cirrus contribution when estimating the kSZ upper limit.
The quoted central values are the medians of each distribution.
The quoted error bars are one half of the 68.3\% confidence interval.
}
\normalsize
\end{center}
\end{table*}

\begin{table*}[ht!]
\begin{center}
\caption{\label{tab:results} Summary of Point Source Results}
\small
\begin{tabular}{cc|ccc|c}
\hline\hline
\rule[-2mm]{0mm}{6mm}
Signal&  &  $150\times150$ ($\mu {\rm K}^2$) & $150\times220$ ($\mu {\rm K}^2$)  & $220\times220$ ($\mu {\rm K}^2$)   & $\bar \alpha_{150-220}$ \\
\hline
\multirow{2}{*}{Poisson }  & $C^{PS,P}$ & 7.36 ($\pm$ 0.59) $\times 10^{-6}$& 2.21 ($\pm$ 0.16) $\times 10^{-5}$& 7.88 ($\pm$ 0.91) $\times 10^{-5}$& \multirow{2}{*}{3.68 ($\pm$ 0.20) } \\
& ${\cal D}^{PS,P}_{3000}$ & 10.5 ($\pm$ 0.8) & 31.7 ($\pm$ 2.3)& 113 ($\pm$ 13) &  \\
(Radio subtracted)  & $C^{PS,P}$ & 6.57 ($\pm$ 0.70) $\times 10^{-6}$& 2.15 ($\pm$ 0.17) $\times 10^{-5}$& 7.84 ($\pm$ 0.91) $\times 10^{-5}$& 3.86 ($\pm$ 0.23 )  \\
\hline
&&&&\\
Clustered & ${\cal D}^{PS,C}_{3000}$  & $< 14$ (95\%) & 16.4 ($\pm$ 4.8) & 54 ($\pm$ 18) & 3.8 ($\pm\,1.3$) \\
\hline
\hline 
 \end{tabular}
 \tablecomments{Spectral indices and power for the Poisson and clustered point sources 
in the $150\times150$, $150\times220$, and $220\times220$ spectra after masking the brightest sources.  
Recall that ${\cal D}_\ell \equiv \ell(\ell+1) C_\ell/(2\pi)$.  
The quoted central values are the medians of each distribution, and the quoted error bars are one half 
of the 68.3\% confidence interval.
The central values for the Poisson and clustered power are from chains with varying Poisson, CMB, clustered, and cirrus contributions and with the fixed fiducial kSZ (${\cal D}_{3000}^{\rm kSZ} = 2.0\, \mu {\rm K}^2$) contribution and the corresponding best-fit tSZ contribution from L10 (${\cal D}_{3000}^{\rm tSZ} = 3.3\, \mu {\rm K}^2$ at $150\times150$).
For the Poisson term, $\bar \alpha_{150-220}^P$ is determined from all three spectra with an assumed upper limit on the variation of the spectral index of $\sigma_\alpha <  0.7$ as discussed in the text.
We also quote the effective power determination and spectral index for the Poisson component after subtracting an estimate of the residual radio source contribution using the \citet{dezotti05} model.
The uncertainties are slightly larger in this case due to the assumed 50\% uncertainty on the subtracted radio source power.
The clustered power upper limit at $150\times150$ assumes no kSZ, tSZ, or cirrus contributions and is quoted at 95\% confidence.
The clustered component spectral index, $\alpha_{150-220}^C$, is determined from the clustered power
determinations for the $150\times220$ and $220\times220$ spectra with the assumption that the fluctuations are completely correlated between bands. 
}
 \normalsize
 \end{center}
\end{table*}

\section{Comparison with Models}
\label{sec:evolmodels}

Here we make comparisons between modeled and observed frequency dependences for both Poisson and clustered power.  We first examine the frequency dependence near the SPT observing frequencies.  Our main conclusion is that the Poisson power favors models with steeper spectral indices.  The steep frequency dependence of our detected Poisson power favors $\beta = 2$ over $\beta = 1.5$ and requires an adjustment to the spectral templates used in the LDP model.

Extending to higher frequencies, we find our single-SED model is able to reproduce the frequency dependence of the clustered power from our 150 and 220~GHz measurements to measurements at $1900\,$GHz.
In fitting across this wide frequency range, we become highly sensitive to a degenerate combination of dust temperature and redshift distribution.  We discuss the redshift distribution of the emissivity density used to match the data, and also compare to the shape of the mean spectrum inferred from {\it COBE}/FIRAS data.   

\subsection{Spectral Indices at SPT Frequencies}
\label{sec:DSFGmodels}

In Table~\ref{tab:results}, we show  the amplitude of the Poisson point source power inferred from each of the three SPT power spectra.  
Assuming that each source has a spectral index between 150 and $220\,$GHz drawn from an independent normal distribution with mean $\bar \alpha_{150-220}^P$ and variance, $\sigma^2_\alpha$, we can reconstruct the values of these two parameters from the Poisson amplitudes.\footnote{Note that 150 and $220\,$GHz are only nominally the frequencies of these two (broad) channels.  
As described in the Appendix, the effective band centers, for the spectral shape expected for dusty galaxies, are 154.2 and $221.3\,$GHz; these are the frequencies we use for relating flux ratios to spectral indices.}  
The constraint on $\sigma^2_\alpha$ from the data is quite weak, and we apply a uniform prior of $0.2 < \sigma_\alpha < 0.7$ to determine $\bar \alpha_{150-220}^P = 3.68 \pm 0.20$.  This prior on $\sigma_\alpha$ is motivated later in this section.

This spectral index result does not include a correction for the subdominant radio source contribution. 
According to the \citet{dezotti05} model, radio sources below our $5\sigma$ masking threshold will account for 11\% of the reported Poisson power at $150\,$GHz and 0.5\% of the Poisson power at $220\,$GHz. 
Subtracting the predicted radio power spectra with an assumed 50\% uncertainty
 steepens the measured spectral index and increases the uncertainty to give 
$\bar \alpha_{150-220}^P = 3.86\pm 0.23$.  

We also consider the impact of the uncertain shape of the ``flat" power components on our inference of $C^{PS,P}$.  
We test the impact by fitting a single flat power component with a range of shapes together with a Poisson term.
For the flat power component, we consider the range of tSZ and kSZ models in L10 in addition to the single-SED model in this work.
The resultant, most likely $C^{PS,P}$ values vary by $3 \times 10^{-7} \mu$K$^2$
which leads to a spread in most likely $\bar \alpha_{150-220}^P$ of 0.07.  This uncertainty has been added in quadrature to the $\bar \alpha_{150-220}^P$ values above and in Table~\ref{tab:results}.

Most models for millimeter-wavelength dust emission predict shallower spectral indices. For example, the Poisson spectral index is 2.3 in the LDP model.  
The LDP model is in good agreement with published source counts at higher frequencies, and it is possible that the disagreement with the SPT data results entirely from the extrapolation to our frequencies.  
All six of the template SEDs in the LDP model have $\alpha_{150-220} \simeq 2.7$ for $z = 2$.  
Expanding the set of spectral templates to include some which are steeper at long wavelengths might resolve the disagreement.

We can also use our measurement of $\bar \alpha_{150-220}^P$ to constrain the dust emissivity index in the single-SED model.  In the RayleighÐJeans (RJ) infinite-temperature limit $\alpha = \beta + 2$, but finite temperature corrections to this
relation are significant.  Taking $T_d = 34\,$K (consistent with \citealt{chapman05} who find $T_d=36 \pm 7\,$K 
and \citealt{dunne00} who find $T_d=35.6 \pm 4.9\,$K), we find $\alpha_{150-220} = \beta + 1.7$ for sources at $z=1$
and $\alpha_{150-220} = \beta + 1.5$ for sources at $z=2$.  

Of course, the emission is coming from a range of redshifts, not just $z=1$ or $z=2$.
If we take the redshift distribution of the Poisson contribution to be the same as that of the mean background light in our fiducial single-SED model, we find $\bar \alpha_{150-220} ^P= 1.38 + \beta$. This implies that $\beta = 2.48 \pm 0.23$.  
Changing $T_d$ from $34\,$K to $40\,$K or changing $z_c$ from $z_c = 2$ to $z_c = 1$, we find $\beta = 2.38 \pm 0.23$.  However, BLAST and {\it Spitzer} data place some strong constraints on these variations as we describe below.  

Theoretical models calibrated with laboratory data and astronomical observations at ultraviolet and visible wavelengths
\citep{draine84} lead to spectra well approximated at long wavelengths with $\beta \sim 2$ \citep{gordon95}.  As
reviewed in \citet{dunne01}, astrophysical observations of dust in different environments are consistent with $\beta$
values between 1 and 2.  Our data clearly prefer the high end of this range.  We note that \citet{meny07} find that
effects due to long-range disorder in the dust grains can lead to $\beta > 2$ at long wavelengths. However, we emphasize that our results are consistent with $\beta=2$ which is within  $2.1\sigma$ of our best-fit value.

Although $\beta \sim 2$ is not surprising from the standpoint of theoretical calculations of the spectral properties of dust, many observations lead to constraints that appear inconsistent with $\beta = 2$.  
For example, \citet{dunne00} find $\beta=1.3 \pm 0.2$ for a sample of 104 galaxies selected at 5$\,$THz (60~$\mu$m), and with fluxes measured at 350$\,$GHz and 3$\,$THz.  And \citet{kovacs06} find from observations of 15 galaxies
that $T_d = 34.6 \pm 3 {\rm K} (1.5/\beta)^{0.71}$ which for $\beta = 2$ is somewhat inconsistent with our assumed
temperature of $34\,$K.  
However, relaxing the assumption of a single temperature can reconcile the data with $\beta = 2$.
For example, \citet{dunne01} find a good fit with $\beta = 2$ by introducing a $20\,$K component to preserve the observed shallow spectral index from 350$\,$GHz to 3$\,$THz.  

\begin{figure*}
\begin{center}
\centerline{
\includegraphics[width=8.5cm]{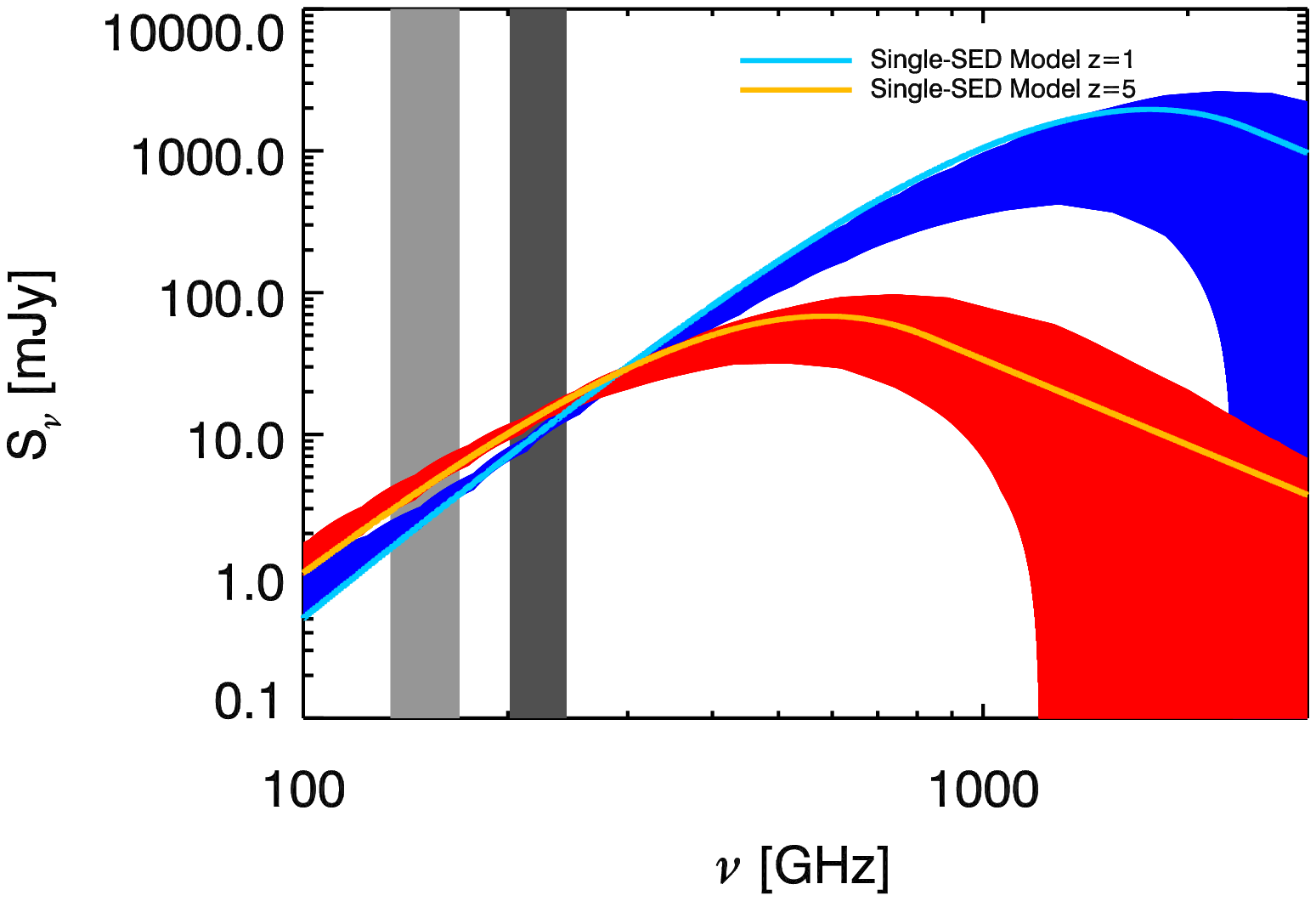}
\includegraphics[width=8.5cm]{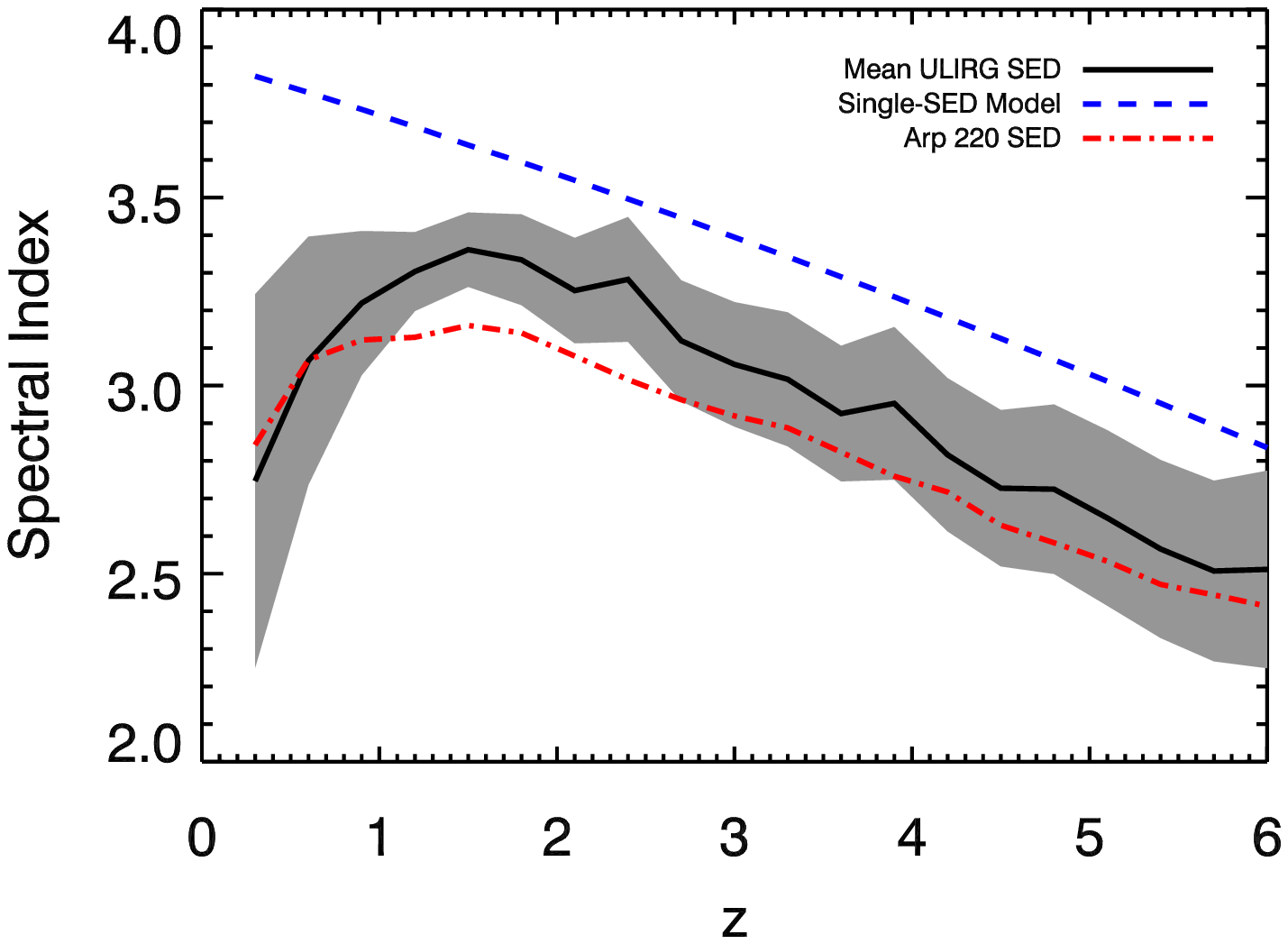}}
\end{center}
\caption{\small {\it Left hand side:} The range of galaxy SED templates from \citet{silva98} at $z=1$ (\emph{blue shaded region}) and $z=5$ (\emph{red shaded region}), normalized to be $20\,$mJy at $220\,$GHz. The SED for the fiducial single-SED model is also shown at $z=1$ (\emph{light blue line}) and $z=5$ (\emph{orange line}).  The shaded vertical regions show the positions and widths of the 150 and 220~GHz SPT bands used in this work.  
{\it Right hand side:} The 67\% interval for 150 to $220\,$GHz model spectral indices at redshifts between 0 and 6 for the \citet{silva98} galaxy SED templates.  The fit to the Arp 220 SED from S98 (\emph{red, dot-dashed line}) and the fiducial single-SED model SED (\emph{blue, dashed line}) are also shown for comparison.
\label{fig:DSFGseds}}
\end{figure*}

We also compare our results with expectations from SEDs for which the spectral properties of the dust are calculated from first principles, and which include synchrotron and free--free emission \citep[hereafter S98]{silva98}.
The S98 templates were constructed from a detailed model of star and dust formation and calibrated with available data.  
In the left panel of Figure~\ref{fig:DSFGseds}, we show the SEDs at $z=1$ and $z=5$ with luminosities set so that their fluxes are $20\,$mJy at the observed $\nu = 220\,$GHz.  The single-SED model is also shown for comparison.  
In the right panel we show, as a function of redshift, the distribution of spectral indices derived from a uniform sampling of the S98 templates and reduced to a mean and $1\sigma$ interval.  
The softening of spectral indices at lower redshifts is due to synchrotron and free--free emission.  Assuming the distribution of template SEDs for galaxies is uniform, and that the redshift distribution of galaxies follows that of the $150\,$GHz background light of our fiducial single-SED model, we find a distribution of spectral indices with mean $\bar \alpha_{150-220} = 3.04$ and $\sigma_\alpha = 0.28$.  
This mean is inconsistent with our observationally determined value of $\bar \alpha_{150-220}^P=3.86\pm0.23$.  This difference is evidence that galactic spectra are preferentially more like the steepest S98 templates than the shallower ones.

In order to constrain the residual DSFG Poisson power in the DSFG-subtracted map, L10 make use of a prior on the scatter in spectral indices of 0.2 $<$ $\sigma_\alpha$ $<$ 0.7.  
Our exercise with the S98 templates would argue for $\sigma_\alpha \simeq 0.3$, but our difference with the S98 models argues for some caution and, therefore, we slightly more than double the dispersion for an upper bound of 0.7.  
Imposing a lower limit on $\sigma_\alpha$ is also conservative when used to place an upper limit on residual power; lower values of $\sigma_\alpha$ will produce less residual power.

The DSFG clustering is subdominant to either the CMB anisotropy or Poisson DSFG power at all angular scales, and is therefore measured less precisely than the Poisson power.  
In addition, we have the difficulty of separating the kSZ contribution which may have a similar shape.  Table~\ref{tab:results} reports clustered component amplitudes with the assumption that kSZ is at our fiducial value.  
As shown in the Appendix, the clustering spectral index of $\bar \alpha_{150-220}^C=3.8 \pm 1.3$ reported in Table~\ref{tab:results} is derived from the $150\times220$ and $220\times220$ bandpowers, and hence is independent of tSZ.  
However, the uncertain kSZ amplitude does introduce considerable extra uncertainty in the clustered DSFG amplitude and spectral index $\bar \alpha_{150-220}^C$.  
If kSZ saturated our upper bound of $13\,\mu {\rm K}^2$, we would find $\bar \alpha_{150-220}^C = 5.9 \pm 2.4$.  Assuming kSZ = 0, we find $\bar \alpha_{150-220}^C = 3.6 \pm 1.2$.

We can also compare these spectral index results with those inferred for the DSFGs in SPT data that were resolved out of the diffuse background light with S/N $>$ 4.5 at both 150 and $220\,$GHz and that do not have {\it IRAS} counterparts. Six of these sources are reported in V09.  
The posterior probability distribution of the spectral index for a source chosen at random from these six is approximately fit by a Gaussian with $\bar \alpha_{150-220} = 3.3 \pm 0.7$. 
The uncertainty includes noise and therefore represents an upper limit on the underlying scatter in the spectral indices of these sources. 
This is in reasonable agreement with the $\bar \alpha_{150-220}^P = 3.86 \pm 0.23$ calculated in this work for the fainter background population.

Finally, we compare with observations made at higher frequencies but of objects with redshifts that are
lower than the redshifts from which the bulk of the CIB light arises.  
These are of interest because they sample
galaxies in the same rest-frame frequency range as probed by the SPT observations of the CIB.   
\citet{dye09} present SED fits assuming $\beta = 2$ for 114 BLAST sources, mostly at $z < 1$, with 1.9 and 4.3$\,$THz {\it Spitzer} 
observations supplementing the BLAST bands.  
\citet{dunne01} find that the {\it IRAS} galaxy fluxes measured at 3 and 5$\,$THz by {\it IRAS} and at 350 and 670$\,$GHz by SCUBA require a two-temperature dust model with $\beta$ between 1.5 and 2.  These observations which probe the same rest-frame 
frequency as the SPT observations, are roughly consistent with the high value of $\beta$ we deduce from the SPT data.  

\subsection{Frequency-dependence of CIB Fluctuation Power}

\begin{figure}[ht]\centering
\includegraphics[scale=0.7]{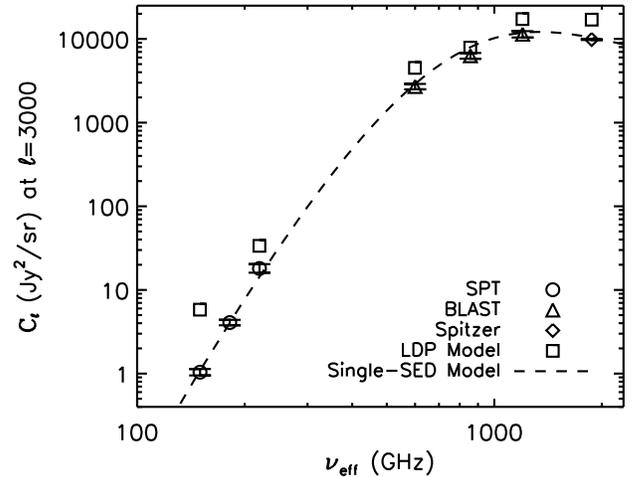}
  \caption[]{\small Amplitude of the Poisson point source power, $C^{PS,P}_{\nu_1\times\nu_2}$, as a function of $\nu_{\rm eff}=\sqrt{\nu_1 \times \nu_2}$ inferred from SPT, BLAST~\citep{viero09}, and {\it Spitzer}~\citep{lagache07} data.  Also shown are the predictions from integrating the LDP model (\emph{squares}) source counts up to the following flux cuts:  $6.4\,$mJy at $150\,$GHz, $17\,$mJy at $220\,$GHz (extrapolation of the $150\,$GHz flux cut with $\alpha_{150-220}=2.7$), no flux cut at $600\,$GHz, $400\,$mJy at $860\,$GHz, $500\,$mJy at $1200\,$GHz, and $200\,$mJy at $1900\,$GHz.  To compare the frequency dependence of the Poisson and clustered power, we also show the single-SED model clustered component frequency dependence (\emph{dashed line}, this is equivalent to the shape of the solid line in Figure~\ref{fig:clust_vs_freq}).} 
\label{fig:poiss_vs_freq}
\end{figure}

\begin{figure}[ht]\centering
\includegraphics[scale=0.7]{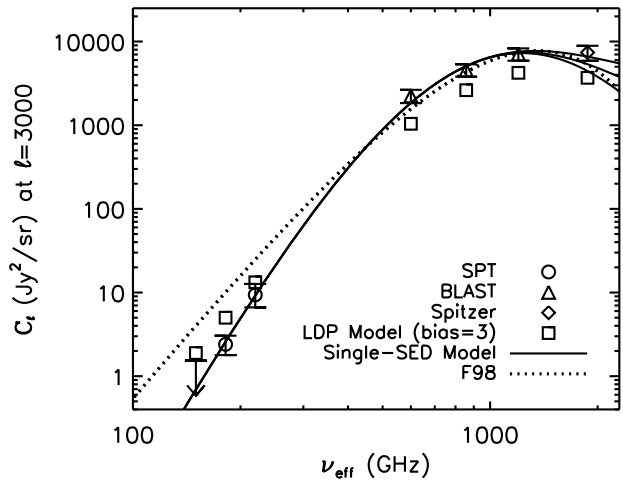}
  \caption[]{\small Amplitude of clustered point source power, $C^{PS,C}_{\nu_1\times\nu_2,l}$ at $\ell = 3000$, as a function of $\nu_{\rm eff}=\sqrt{\nu_1 \times \nu_2}$ inferred from SPT, BLAST \citep{viero09}, and {\it Spitzer} 1.9$\,$THz data \citep{lagache07}.  At $150\,$GHz the 95\% confidence upper limit on the amplitude is given (\emph{down arrow}).  Also shown are predictions of the LDP model with bias set to 3 (\emph{squares}) and the same flux cuts as in Figure~\ref{fig:poiss_vs_freq}.  Our single-SED model is shown (\emph{solid line}) with amplitude and parameters chosen to match SPT and BLAST power ($T_d=34$\,K, $z_c=2$, $\sigma_z=2$, $\beta=2$ and the following three values of $\alpha_{\rm mid-IR}$ in order of decreasing power for comparison: 0.5, 1, and 2).  The mean inferred by \citet{fixsen98} from {\it COBE}-FIRAS data is also shown (\emph{dotted curve}), which has a spectral dependence of $I_\nu \propto \nu^{0.64}B_\nu(18.5\,{\rm K})$ where $B_\nu$ is the Planck function.  All inferences here (except for the upper limit) assume our fiducial value of kSZ is correct at 2.0 $\,\mu {\rm K}^2$.}
\label{fig:clust_vs_freq}
\end{figure}

In Figure~\ref{fig:poiss_vs_freq}, we show the frequency dependence of Poisson power (in Jy$^2$ sr$^{-1}$) as inferred from SPT data at 150 and 220$\,$GHz, BLAST data at 600, 860, and 1200$\,$GHz \citep{viero09}, and {\it Spitzer} data at 1900$\,$GHz \citep{lagache07}. In Figure~\ref{fig:clust_vs_freq}, we do the same for the clustered power.  
The SPT data points are determined under the assumption of the fiducial kSZ amplitude.   We find that we can choose parameters for our single-SED model that reproduce the observed frequency dependence of the clustered power.

Since the shapes of graybody spectra from individual objects at redshift $z$ with temperature $T$ are identical for
equivalent $T/(1+z)$, we have a degeneracy between the temperature and redshift distribution.  
Thus, we fix the temperature when exploring the implications of our data in the context of the single-SED model.  
If the average temperature is higher than we assume, then the redshift distribution will be shifted toward lower redshifts than what we infer.  

For the parameter values from Section~\ref{sec:DSFGmodels} ($\beta = 2$ and $T_d = 34$~K), we find a good match to the data by setting $z_c=2$ and $\sigma_z=2$.  With these redshift-distribution parameters, the peak of the $150\,$GHz $d\bar I_\nu/dz$ is at $z = 3.1$, and half the $150\,$GHz background light arises from $z > 3.2$.

However, this inferred redshift distribution critically depends on the assumption of a single dust temperature of $T_d = 34\,$K.  We know that the spectra of some galaxies require significant additional cold components.  Were we to include an additional (colder) dust component, we would find the redshift distribution shifted toward lower redshifts.  Light from $z=2.4$ and $T_d= 34\,$K is spectrally the same as light from $z=1$ and $T_d=20\,$K.  We do not include multiple dust temperatures in our modeling because the small number of data points does not warrant such complexity.

We can compare the $d\bar I_\nu/dz$ of our model with that of  the LDP model as a  check on our consistency with known counts.  We find that the mean redshifts of the CIB light in the LDP model and our single-SED model are quite similar:  at 150, 600, and $1200\,$GHz the LDP model mean redshifts are 3.2, 2.2, and 1.6, respectively, and the fiducial single-SED model mean redshifts are 3.3, 2.7, and 2.1.  We can also compare our model with \citet{marsden09} who find the fraction of background light at $z > 1.2$ at 600, 860, and 1200$\,$GHz to be 61\%, 51\%, and 40\%, respectively.  This result relies on a stacking analysis using the FIDEL survey galaxies as a very deep catalog of sources of CIB flux.  Uncertainties in their result arise from possible incompleteness and from errors in the photometrically-determined division of FIDEL galaxies into $z<1.2$ and $z>1.2$ subcatalogs.  In our fiducial single-SED model, there is significantly more light at $z > 1.2$ at these frequencies:  95\%, 91\%, and 85\%, respectively.  These differences may be due to the above-mentioned uncertainties in the \citet{marsden09} result or they may be accommodated by a redshift distribution with more freedom than our two parameters allow, multiple SEDs, or multiple temperatures as discussed above. 
 
Our single SED has the exponential decline on the Wien side of the graybody replaced by a power-law decline $\nu^{-\alpha_{\rm mid-IR}}$ as described and motivated in Section~\ref{subsec:clustering}.   
Adjusting $\alpha_{\rm mid-IR}$ only affects the curve near the {\it Spitzer} data point in Figure~\ref{fig:clust_vs_freq}; values $\alpha_{\rm mid-IR} = 0.5$, $1$, and $2$ (in order of decreasing power) are shown.

As noted above, our single-SED model is not detailed enough to make predictions for Poisson power.  However, we find that the observed Poisson power has a frequency dependence that is very similar to that of the clustered power.  This similarity indicates that the spectral properties do not strongly depend on galaxy brightness, redshift, and bias.  

We can use our fiducial model above to relate $\bar \alpha_{150-220}^P$ to $\beta$.  Allowing higher $T_d$ or a distribution
shifted toward lower redshift decreases $\beta$ for fixed $\bar \alpha_{150-220}^P$.  However, the BLAST and {\it Spitzer} clustered power measurements limit our ability to relax the constraints on $\beta$ in this way.  By adjusting the temperature and the redshift distribution, we can at best reduce $\beta$ by 0.03 and maintain the same quality
of fit.  

We find that a linear-theory bias of $\sim 4$ is required for the LDP model to match the BLAST power, in agreement with \citet{viero09}.  Such a bias would overpredict SPT power.  The discrepancy could potentially be resolved by a frequency-dependent bias \citep{knox01}.  But the discrepancy between modeled and observed frequency dependence of the clustered power is similar to what we see with the Poisson power, and the Poisson power is unaffected by bias.   Model adjustments to address the (more severe) Poisson power frequency-dependence discrepancy (such as the change to SEDs we suggest in this section) would, at the very least, go a long way toward eliminating the clustered power frequency-dependence discrepancy as well.  Recall that the linear-theory bias is not a halo-model bias.  Using a halo-model extension of the LDP model, \citet{viero09} find $b \simeq 2.4$.

\citet{fixsen98} used {\it COBE}-FIRAS data to estimate the mean of the background light from 150 to $2400\,$GHz.  Their best fit is $\bar I_\nu = (\nu/\nu_0)^\beta B_\nu(18.5\pm1.2\,{\rm K})$ with $\beta = 0.64\pm0.12$.  This form is to be interpreted phenomenologically; i.e., it is not meant to represent emission from an object with a single temperature.  It can be understood as resulting from the sum of a number of graybody spectra with differing $T/(1+z)$ and higher $\beta$.  The spread in $T/(1+z)$ values softens the rise in brightness with increasing frequency, since as frequency increases more objects are being sampled on the Wien side.
The steeper rise we infer from SPT data does not contradict the FIRAS data which have no constraining power at $220\,$GHz and lower frequencies.  

\subsection{kSZ Models}

The contribution to kSZ from patchy reionization is dependent on the details of the transition from neutral to ionized inter-galactic medium and is thus highly uncertain.  
Our upper limit on kSZ is stringent enough to rule out the very highest predictions in the literature, which range as high as 15 $\,\mu {\rm K}^2$ at $\ell = 3000$ \citep{santos03}\footnote{The power is for a model with $\tau = 0.11$; the predictions range even higher if we allow higher values of $\tau$.}.  
In more recent models (e.g., \citealt{zahn05}), based on excursion set analytic models and full radiative transfer simulations, the patchy reionization signal is largest on scales of $\ell \simeq 2000$ where there is still considerable primary anisotropy power, and it is only tens of percents of the total kSZ signal on scales beyond the primaries. 
The modeling of the post-reionization kSZ has also been inconclusive, with amplitudes varying by a factor of $2$. 
Given that we only marginally detect the thermal SZ effect which is generally expected to be larger than the kSZ signal, the current data quality is insufficient to place 
a strong constraint on the amplitude and shape of the kSZ power spectrum. 
The increased sky coverage and addition of a $95\,$GHz channel in the 2009 SPT data will make it possible to isolate and 
exploit this new cosmological probe. 

\section{Residual clustered power in the ``DSFG-subtracted map''}

\label{sec:DSFGjust}

Our $150\,$GHz maps have the following signal components
\begin{eqnarray}
\label{eq:mapcomponents}
\delta T_{150} &=&\delta T_{150}^{PS,C} + \delta T_{150}^{PS,P} + \delta T_{150}^{tSZ} + \nonumber \\
	&&\delta T_{150}^{kSZ} + \delta T_{150}^{CMB} + \delta T_{150}^{\rm Cirrus},
\end{eqnarray}
where $PS,C$ stands for the clustered point source component and $PS,P$ for the Poisson point source component.  
The $220\,$GHz maps are the same with the exception of no tSZ contribution.  The DSFG-subtracted map of L10 is related to the individual
frequency maps by
\be
\label{eq:subtractedmap}
\delta T_s = \left(\delta T_{150} - x \delta T_{220}\right) 
\ee
with $x=0.325$, chosen to minimize the Poisson point source power. This value of $x$ corresponds to
 $\bar \alpha_{150-220} = 3.6$ and is consistent with the spectral indices found in this work.  
 A $1/(1-x)$
normalization factor can be included so that the subtraction has no effect on
CMB temperature fluctuations, but we omit it here so that the subtraction
has no effect on the tSZ signal.  L10 analyze the power spectrum of this map,
assuming it has negligible contribution from clustered point sources, cirrus, and
a constrained amount of residual Poisson power.  
Here, we justify the assumption of negligible clustered power.

From the definition of the DSFG-subtracted map, the clustered power it contains is related to the auto and
cross-clustered power via
\be
\label{eq:modelsubtractedpower}
  {\cal D}^{PS,C}_{s,l}  = \left[{\cal D}^{PS,C}_{150\times150,l} - 2x {\cal D}^{PS,C}_{150\times220,l} + x^2 {\cal D}^{PS,C}_{220\times220,l}\right].
\ee
We do not measure the terms on the right-hand
side well enough to use our measurements alone to provide a useful upper limit
on the level of contamination.  Fortunately, we have additional information to use.

We assume that the spatial fluctuations at $150\,$GHz are 100\% correlated with the fluctuations at $220\,$GHz\footnote{By which we mean $\langle \delta T_{150} \delta T_{220} \rangle =  \sqrt{\langle \delta T_{150}^2 \rangle \langle \delta T_{220}^2 \rangle}$.}.  
For a power-law SED, the 150 and $220\,$GHz window functions would be identical, and we would have perfect correlation.  
Even for galaxies at $z=3$, the SPT observing bands are far on the RJ side of the spectrum and a power-law spectrum is a good approximation.  
Therefore, perfect correlation between bands is a very good approximation \citep{knox01}.  

With this assumption, we have two estimates for the residual $150\,$GHz clustered power:
\be
\label{eq:modelsubtractedpower2}
 {\cal D}^{PS,C}_{s,l}  = {\cal D}^{PS,C}_{220\times220,l}\left(x - x_c\right)^2 
\ee 
and
\be
\label{eq:modelsubtractedpower3}
 {\cal D}^{PS,C}_{s,l}  = {\cal D}^{PS,C}_{150\times220,l}  \left(x - x_c\right)^2 / x_c,
\ee
where $x_c^2$ is the true ratio between the $150\times150$ power and $220\times220$ clustered
power, and $x$ is the assumed value of 0.325. 
 $x_c$ is a function of the spectral index $\alpha_{150-220}^C$ of the clustering component, which we assume (for the moment) to be equal to the spectral index of the Poisson term $\bar \alpha_{150-220}^P$. We approximate the distribution of spectral indices by a normal distribution with $\bar \alpha= 3.86$ and $\sigma_{\alpha}= 0.23$.  

We calculate the likelihood function for ${\cal D}^{PS,C}_{s,l}$ from the $150\times220$ and $220\times220$ data.  Assuming each estimate is independent, we combine the two likelihood functions to produce a combined estimate of the likelihood function according to
 \be
 P( {\cal D}^{PS,C}_{s,l}) \propto  P( {\cal D}^{PS,C}_{s,l} | 150\times220) P( {\cal D}^{PS,C}_{s,l} | 220\times220)
 \ee
 and integrate it to find the 95\% confidence upper limit: ${\cal D}^{PS,C}_{s,l} < 0.3\,\mu {\rm K}^2$ at $\ell = 3000$. 
This result has very little dependence on kSZ.  This level of clustered point source power is small
compared to the tSZ power.  If we instead use the radio-unsubtracted spectral index of $3.68 \pm 0.20$ to approximate the $x_c$ distribution we get a lower 95\% confidence upper limit of ${\cal D}^{PS,C}_{s,l} < 0.1\,\mu {\rm K}^2$ at $\ell = 3000$. 

We now consider how much $\alpha_{150-220}^C$ might differ from $\bar \alpha_{150-220}^P$.  We expect these two indices to be similar due to the fact that these are describing the spectra of light from the same objects.  However, the light contributing to the determination of these indices comes from different weightings of these same objects with the Poisson power being more heavily weighted to brighter objects.  One source of difference then is the dependence of the redshift distribution on flux.  To quantify how much different redshift distributions could affect the difference between $\alpha_{150-220}^C$ and $\bar \alpha_{150-220}^P$, we turn to the \citet{righi08} model, which has very large amounts of the light coming from faint objects at high redshift.  For this extreme model, the resulting clustered and Poisson indices shift with $\alpha_{150-220}^P - {\bar \alpha_{150-220}^C} = 0.2$.  We thus conservatively test the effect of shifting the clustered power spectral index with respect to the Poisson spectral index by as much as 0.4, double the difference in the \citet{righi08} model.  We repeat the upper limit calculation of the last paragraph with a mean spectral index of 3.46 instead of 3.86 with the uncertainty unchanged.   With this distribution, we find a stronger upper limit of $0.04\,\mu {\rm K}^2$.  Since we have very little direct empirical knowledge of the sources in the faint magnitude ranges that make up the bulk of the CIB light at millimeter wavelengths, we adopt the original, more conservative assumption to quantify our uncertainty about residual CIB clustered power in the L10 DSFG-subtracted map.

\section{Conclusions}
\label{sec:conclusions}

We present detections of both Poisson and clustered power from DSFGs from an analysis of the SPT angular temperature anisotropy bandpowers presented in a companion paper (L10). 
This is the first detection of the DSFG clustering component at millimeter wavelengths. 
We find the Poisson power at $\ell = 3000$ to be $10.5\pm0.8, 31.7\pm2.3, 113\pm13 \,\mu {\rm K}^2$ for the $150\times150$~GHz, $150\times220$~GHz, and $220\times220$~GHz power spectra, respectively. 
We find the clustered power amplitude at $\ell = 3000$ to be approximately one half as large, with values of  $16.4\pm4.8$ and $54\pm18\,\mu {\rm K}^2$ at $150\times220$ and $220\times220$ respectively. 
At $150\,$GHz, where there is a degeneracy between the tSZ effect, kSZ effect, and clustered DSFG component, we set an upper limit on the clustered power of 14$\,\mu {\rm K}^2$ at 95\% confidence. 
These probes of the statistical properties of the DSFG population complement the SPT source catalog presented by V09 which measures the bright tail of the DSFG and AGN flux distributions.

We also use the $150\,$GHz bandpowers to constrain the amplitude of the kSZ power spectrum to be $\leq 13 \,\mu {\rm K}^2$ at $\ell = 3000$ with 95\% confidence. 
This upper limit begins to rule out the highest of kSZ models which reach 15$\,\mu {\rm K}^2$ for $\tau = 0.11$ \citep{santos03}.  
With more SPT data, including that from the $95\,$GHz band, we can expect tighter constraints on the kSZ effect 
and perhaps new information about the reionization history of the universe.

We find the spectral dependence of the Poisson power due to DSFGs between 150 and $220\,$GHz to be $\bar\alpha_{150-220}^{P} = 3.86 \pm 0.23$ which is on the steeper end of the expected range and higher than many current models.  However, our results are consistent with theoretical expectations for the spectral properties of interstellar dust and, under the assumption of no evolution of those spectral properties, with low-redshift observations of individual galaxies at the relevant wavelengths.  

Combining the SPT results with those from BLAST and {\it Spitzer}, we show the frequency dependence of the Poisson and clustered power from $150\,$GHz to almost $2000\,$GHz ($160\, \mu$m).
The clustered power over this large frequency range is consistent with a simple ``single-SED'' model of graybody spectra with a temperature of $T_d = 34$\,K, $\beta = 2$, and a broad redshift distribution of the emissivity density, with half the light at $150\,$GHz coming from $z > 3.2$. 
We do not claim this model is a unique description of the current data,
but over this broad range of frequencies the data tightly constrain the combination of dust temperature and redshift distribution.  The dust temperature of 34\,K is consistent with \citet{dunne00} and \citet{chapman05}.  
The redshift distribution is similar to that of the LDP model, but shifted to higher redshifts than the distribution inferred by \citet{marsden09}.  We find it difficult to accommodate values of $\beta$ less than 2; reconciling this steep index with measurements of individual galaxies at shorter wavelengths may require dust with multiple temperatures.  
 
In a companion paper, SPT data are used by V09 to detect the brightest members of the source populations that are likely to contribute to these Poisson and clustered signals.   
Using the flux ratio between 150 and $220\,$GHz, these sources are separated into those with spectra consistent with active galactic nuclei (AGN) and those consistent with DSFGs.  
V09 present source counts for each population and argue that most of the detected DSFGs, having no counterparts in the {\it IRAS} catalog, must reside at high redshift or have unphysically cold dust.  

Furthermore, we argue that it is safe to neglect the clustering of radio sources given extrapolations to $150\,$GHz of the clustered radio power from observations of the mean temperature (suggesting less than  $\sim 0.25\, \mu {\rm K}^2$) and from source counts (suggesting less than 0.01$\,\mu {\rm K}^2$).  
Several lines of reasoning also suggest that the correlations of radio galaxies and DSFGs with the SZ signal due to galaxy clusters will be negligible.

L10 derive constraints on SZ power from a DSFG-subtracted map constructed from a linear combination of the 150 and $220\,$GHz maps. 
A residual clustered DSFG component would bias the SZ constraints presented by L10. 
The residual DSFG power can be predicted from the average spectral index of the DSFG population, the measured clustered power amplitudes at $150\times220$ and $220\times220$, and the exact frequency combination used to construct the DSFG-subtracted bandpowers. 
We assume that the frequency dependence of the clustered component is the same as the Poisson component.
With this assumption, we place an upper limit  at 95\% confidence
on the clustered component in the DSFG-subtracted band powers of $0.3\,\mu {\rm K}^2$ at $\ell = 3000$.
We also argue that the variation of spectral indices between sources is expected to be small; $\sigma_\alpha \in [0.2, 0.7]$.
This constraint on spectral index variability is converted into a prior on the residual Poisson DSFG component in the DSFG-subtracted maps in L10.

A full multifrequency analysis is a natural progression from the single-frequency bandpower parameter fits performed in this work, and we are in the process of developing a simultaneous multifrequency fitting pipeline. 
In addition, the SPT bandpowers used in this work represent only a small fraction of the complete SPT data set.  
The complete SPT survey will cover over $2000\, {\rm deg}^2$ at 95, 150, and $220\,$GHz. 
The three frequencies will allow us to separate the kSZ from DSFGs and tSZ and place new constraints on the reionization 
history of the universe.  

\acknowledgments

We thank Marco Viero and Guilaine Lagache for comparison of calculations as well as Andrew Blain, Douglas Scott, Rashid Sunyaev, Simon White, and George Efstathiou for useful conversations.  The SPT team gratefully acknowledges the contributions to the design and construction of the telescope by S.\ Busetti, E.\ Chauvin,  T.\ Hughes, P.\ Huntley, and E.\ Nichols and his team of iron workers. We also thank the 
National Science Foundation (NSF) Office of Polar Programs, the United States Antarctic Program, and the Raytheon Polar Services Company for their support of the project.  We are grateful for professional support from the staff of the South Pole station. 
We thank T.\ Lanting, J.\ Leong, A.\ Loehr, W.\ Lu, M.\ Runyan, D.\ Schwan, M.\ Sharp, and C.\ Greer for their early contributions to the SPT project and J.\ Joseph and C.\ Vu for their contributions to the electronics. 

The South Pole Telescope is supported by the National Science Foundation through grants ANT-0638937 and ANT-0130612.  Partial support is also provided by the  
NSF Physics Frontier Center grant PHY-0114422 to the Kavli Institute of Cosmological Physics at the University of Chicago, the Kavli Foundation, and the Gordon and Betty Moore Foundation. 
This research used resources of the National Energy Research Scientific Computing Center, which is supported by the Office of Science of the U.S. Department of Energy under Contract No. DE-AC02-05CH11231.
The McGill group acknowledges funding from the National Sciences and
Engineering Research Council of Canada, the Quebec Fonds de recherche
sur la nature et les technologies, and the Canadian Institute for
Advanced Research. The following individuals acknowledge additional support:
J.\ McMahon from a Fermi Fellowship,  Z.\ Staniszewski from a GAAN Fellowship, A.T.\ Lee from the Miller Institute for Basic Research in Science, University of California Berkeley, N.W.\ Halverson from an Alfred P.\ Sloan Research Fellowship, and K.\ Schaffer, B.A.\ Benson, and E.R.\ Switzer from KICP Fellowships.

\bibliography{spt_smg,../../BIBTEX/spt}

\begin{thebibliography}{67}
\expandafter\ifx\csname natexlab\endcsname\relax\def\natexlab#1{#1}\fi

\bibitem[{{Bai} {et~al.}(2007){Bai}, {Marcillac}, {Rieke}, {Rieke}, {Tran},
  {Hinz}, {Rudnick}, {Kelly}, \& {Blaylock}}]{bai07}
{Bai}, L., {Marcillac}, D., {Rieke}, G.~H., {Rieke}, M.~J., {Tran}, K., {Hinz},
  J.~L., {Rudnick}, G., {Kelly}, D.~M., \& {Blaylock}, M. 2007, \apj, 664, 181

\bibitem[{{Blain} {et~al.}(2003){Blain}, {Barnard}, \& {Chapman}}]{blain03}
{Blain}, A.~W., {Barnard}, V.~E., \& {Chapman}, S.~C. 2003, \mnras, 338, 733

\bibitem[{{Blain} {et~al.}(1999){Blain}, {Kneib}, {Ivison}, \&
  {Smail}}]{blain99}
{Blain}, A.~W., {Kneib}, J.~., {Ivison}, R.~J., \& {Smail}, I. 1999, \apjl,
  512, L87

\bibitem[{{Blain} {et~al.}(2002){Blain}, {Smail}, {Ivison}, {Kneib}, \&
  {Frayer}}]{blain02}
{Blain}, A.~W., {Smail}, I., {Ivison}, R.~J., {Kneib}, J.-P., \& {Frayer},
  D.~T. 2002, \physrep, 369, 111

\bibitem[{{Bond} {et~al.}(1986){Bond}, {Carr}, \& {Hogan}}]{bond86}
{Bond}, J.~R., {Carr}, B.~J., \& {Hogan}, C.~J. 1986, \apj, 306, 428

\bibitem[{{Bond} {et~al.}(1991){Bond}, {Carr}, \& {Hogan}}]{bond91b}
---. 1991, \apj, 367, 420

\bibitem[{{Chapman} {et~al.}(2005){Chapman}, {Blain}, {Smail}, \&
  {Ivison}}]{chapman05}
{Chapman}, S.~C., {Blain}, A.~W., {Smail}, I., \& {Ivison}, R.~J. 2005, \apj,
  622, 772

\bibitem[{{Condon}(1974)}]{condon74}
{Condon}, J.~J. 1974, \apj, 188, 279

\bibitem[{{de Zotti} {et~al.}(2005){de Zotti}, {Ricci}, {Mesa}, {Silva},
  {Mazzotta}, {Toffolatti}, \& {Gonz{\'a}lez-Nuevo}}]{dezotti05}
{de Zotti}, G., {Ricci}, R., {Mesa}, D., {Silva}, L., {Mazzotta}, P.,
  {Toffolatti}, L., \& {Gonz{\'a}lez-Nuevo}, J. 2005, \aap, 431, 893

\bibitem[{{Dole} {et~al.}(2006){Dole}, {Lagache}, {Puget}, {Caputi},
  {Fern{\'a}ndez-Conde}, {Le Floc'h}, {Papovich}, {P{\'e}rez-Gonz{\'a}lez},
  {Rieke}, \& {Blaylock}}]{dole06}
{Dole}, H., {Lagache}, G., {Puget}, J.-L., {Caputi}, K.~I.,
  {Fern{\'a}ndez-Conde}, N., {Le Floc'h}, E., {Papovich}, C.,
  {P{\'e}rez-Gonz{\'a}lez}, P.~G., {Rieke}, G.~H., \& {Blaylock}, M. 2006,
  \aap, 451, 417

\bibitem[{{Dole} {et~al.}(2004){Dole}, {Le Floc'h}, {P{\'e}rez-Gonz{\'a}lez},
  {Papovich}, {Egami}, {Lagache}, {Alonso-Herrero}, {Engelbracht}, {Gordon},
  {Hines}, {Krause}, {Misselt}, {Morrison}, {Rieke}, {Rieke}, {Rigby}, {Young},
  {Bai}, {Blaylock}, {Neugebauer}, {Beichman}, {Frayer}, {Mould}, \&
  {Richards}}]{dole04}
{Dole}, H., {Le Floc'h}, E., {P{\'e}rez-Gonz{\'a}lez}, P.~G., {Papovich}, C.,
  {Egami}, E., {Lagache}, G., {Alonso-Herrero}, A., {Engelbracht}, C.~W.,
  {Gordon}, K.~D., {Hines}, D.~C., {Krause}, O., {Misselt}, K.~A., {Morrison},
  J.~E., {Rieke}, G.~H., {Rieke}, M.~J., {Rigby}, J.~R., {Young}, E.~T., {Bai},
  L., {Blaylock}, M., {Neugebauer}, G., {Beichman}, C.~A., {Frayer}, D.~T.,
  {Mould}, J.~R., \& {Richards}, P.~L. 2004, \apjs, 154, 87

\bibitem[{{Draine} \& {Lee}(1984)}]{draine84}
{Draine}, B.~T. \& {Lee}, H.~M. 1984, \apj, 285, 89

\bibitem[{{Dunne} {et~al.}(2000){Dunne}, {Eales}, {Edmunds}, {Ivison},
  {Alexander}, \& {Clements}}]{dunne00}
{Dunne}, L., {Eales}, S., {Edmunds}, M., {Ivison}, R., {Alexander}, P., \&
  {Clements}, D.~L. 2000, \mnras, 315, 115

\bibitem[{{Dunne} \& {Eales}(2001)}]{dunne01}
{Dunne}, L. \& {Eales}, S.~A. 2001, \mnras, 327, 697

\bibitem[{{Dwek} \& {Arendt}(1998)}]{dwek98}
{Dwek}, E. \& {Arendt}, R.~G. 1998, \apjl, 508, L9

\bibitem[{{Dye} {et~al.}(2009){Dye}, {Ade}, {Bock}, {Chapin}, {Devlin},
  {Dunlop}, {Eales}, {Griffin}, {Gundersen}, {Halpern}, {Hargrave}, {Hughes},
  {Klein}, {Magnelli}, {Marsden}, {Mauskopf}, {Moncelsi}, {Netterfield},
  {Olmi}, {Pascale}, {Patanchon}, {Rex}, {Scott}, {Semisch}, {Targett},
  {Thomas}, {Truch}, {Tucker}, {Tucker}, {Viero}, \& {Wiebe}}]{dye09}
{Dye}, S., {Ade}, P.~A.~R., {Bock}, J.~J., {Chapin}, E.~L., {Devlin}, M.~J.,
  {Dunlop}, J.~S., {Eales}, S.~A., {Griffin}, M., {Gundersen}, J.~O.,
  {Halpern}, M., {Hargrave}, P.~C., {Hughes}, D.~H., {Klein}, J., {Magnelli},
  B., {Marsden}, G., {Mauskopf}, P., {Moncelsi}, L., {Netterfield}, C.~B.,
  {Olmi}, L., {Pascale}, E., {Patanchon}, G., {Rex}, M., {Scott}, D.,
  {Semisch}, C., {Targett}, T., {Thomas}, N., {Truch}, M.~D.~P., {Tucker}, C.,
  {Tucker}, G.~S., {Viero}, M.~P., \& {Wiebe}, D.~V. 2009, \apj, 703, 285

\bibitem[{{Eales} {et~al.}(1999){Eales}, {Lilly}, {Gear}, {Dunne}, {Bond},
  {Hammer}, {Le F{\`e}vre}, \& {Crampton}}]{eales99}
{Eales}, S., {Lilly}, S., {Gear}, W., {Dunne}, L., {Bond}, J.~R., {Hammer}, F.,
  {Le F{\`e}vre}, O., \& {Crampton}, D. 1999, \apj, 515, 518

\bibitem[{{Fernandez-Conde} {et~al.}(2008){Fernandez-Conde}, {Lagache},
  {Puget}, \& {Dole}}]{fernandez-conde08}
{Fernandez-Conde}, N., {Lagache}, G., {Puget}, J., \& {Dole}, H. 2008, \aap,
  481, 885

\bibitem[{{Finkbeiner} {et~al.}(1999){Finkbeiner}, {Davis}, \&
  {Schlegel}}]{finkbeiner99}
{Finkbeiner}, D.~P., {Davis}, M., \& {Schlegel}, D.~J. 1999, \apj, 524, 867

\bibitem[{{Fixsen} {et~al.}(1998){Fixsen}, {Dwek}, {Mather}, {Bennett}, \&
  {Shafer}}]{fixsen98}
{Fixsen}, D.~J., {Dwek}, E., {Mather}, J.~C., {Bennett}, C.~L., \& {Shafer},
  R.~A. 1998, \apj, 508, 123

\bibitem[{{Fixsen} {et~al.}(2009){Fixsen}, {Kogut}, {Levin}, {Limon}, {Lubin},
  {Mirel}, {Seiffert}, {Singal}, {Wollack}, {Villela}, \&
  {Wuensche}}]{fixsen09}
{Fixsen}, D.~J., {Kogut}, A., {Levin}, S., {Limon}, M., {Lubin}, P., {Mirel},
  P., {Seiffert}, M., {Singal}, J., {Wollack}, E., {Villela}, T., \&
  {Wuensche}, C.~A. 2009, submitted to \apj, astro-ph/0901.0555

\bibitem[{{Gervasi} {et~al.}(2008){Gervasi}, {Tartari}, {Zannoni}, {Boella}, \&
  {Sironi}}]{gervasi08}
{Gervasi}, M., {Tartari}, A., {Zannoni}, M., {Boella}, G., \& {Sironi}, G.
  2008, \apj, 682, 223

\bibitem[{{Gordon}(1995)}]{gordon95}
{Gordon}, M.~A. 1995, \aap, 301, 853

\bibitem[{{Granato} {et~al.}(2004){Granato}, {De Zotti}, {Silva}, {Bressan}, \&
  {Danese}}]{granato04}
{Granato}, G.~L., {De Zotti}, G., {Silva}, L., {Bressan}, A., \& {Danese}, L.
  2004, \apj, 600, 580

\bibitem[{{Grossan} \& {Smoot}(2007)}]{grossan07}
{Grossan}, B. \& {Smoot}, G.~F. 2007, \aap, 474, 731

\bibitem[{{Gruzinov} \& {Hu}(1998)}]{gruzinov98}
{Gruzinov}, A. \& {Hu}, W. 1998, \apj, 508, 435

\bibitem[{{Haiman} \& {Knox}(2000)}]{haiman00}
{Haiman}, Z. \& {Knox}, L. 2000, \apj, 530, 124

\bibitem[{{Hivon} {et~al.}(2002){Hivon}, {G{\' o}rski}, {Netterfield}, {Crill},
  {Prunet}, \& {Hansen}}]{hivon02}
{Hivon}, E., {G{\' o}rski}, K.~M., {Netterfield}, C.~B., {Crill}, B.~P.,
  {Prunet}, S., \& {Hansen}, F. 2002, \apj, 567, 2

\bibitem[{{Hughes} {et~al.}(1998){Hughes}, {Serjeant}, {Dunlop},
  {Rowan-Robinson}, {Blain}, {Mann}, {Ivison}, {Peacock}, {Efstathiou}, {Gear},
  {Oliver}, {Lawrence}, {Longair}, {Goldschmidt}, \& {Jenness}}]{hughes98c}
{Hughes}, D.~H., {Serjeant}, S., {Dunlop}, J., {Rowan-Robinson}, M., {Blain},
  A., {Mann}, R.~G., {Ivison}, R., {Peacock}, J., {Efstathiou}, A., {Gear}, W.,
  {Oliver}, S., {Lawrence}, A., {Longair}, M., {Goldschmidt}, P., \& {Jenness},
  T. 1998, \nat, 394, 241

\bibitem[{{Knox} {et~al.}(2001){Knox}, {Cooray}, {Eisenstein}, \&
  {Haiman}}]{knox01}
{Knox}, L., {Cooray}, A., {Eisenstein}, D., \& {Haiman}, Z. 2001, \apj, 550, 7

\bibitem[{{Knox} {et~al.}(1998){Knox}, {Scoccimarro}, \& {Dodelson}}]{knox98}
{Knox}, L., {Scoccimarro}, R., \& {Dodelson}, S. 1998, Physical Review Letters,
  81, 2004

\bibitem[{{Knudsen} {et~al.}(2008){Knudsen}, {van der Werf}, \&
  {Kneib}}]{knudsen08}
{Knudsen}, K.~K., {van der Werf}, P.~P., \& {Kneib}, J. 2008, \mnras, 384, 1611

\bibitem[{{Kov{\'a}cs} {et~al.}(2006){Kov{\'a}cs}, {Chapman}, {Dowell},
  {Blain}, {Ivison}, {Smail}, \& {Phillips}}]{kovacs06}
{Kov{\'a}cs}, A., {Chapman}, S.~C., {Dowell}, C.~D., {Blain}, A.~W., {Ivison},
  R.~J., {Smail}, I., \& {Phillips}, T.~G. 2006, \apj, 650, 592

\bibitem[{{Lagache} {et~al.}(2007){Lagache}, {Bavouzet}, {Fernandez-Conde},
  {Ponthieu}, {Rodet}, {Dole}, {Miville-Desch{\^e}nes}, \& {Puget}}]{lagache07}
{Lagache}, G., {Bavouzet}, N., {Fernandez-Conde}, N., {Ponthieu}, N., {Rodet},
  T., {Dole}, H., {Miville-Desch{\^e}nes}, M.-A., \& {Puget}, J.-L. 2007,
  \apjl, 665, L89

\bibitem[{{Lagache} {et~al.}(2003){Lagache}, {Dole}, \& {Puget}}]{lagache03}
{Lagache}, G., {Dole}, H., \& {Puget}, J. 2003, \mnras, 338, 555

\bibitem[{{Lagache} {et~al.}(2004){Lagache}, {Dole}, {Puget},
  {P{\'e}rez-Gonz{\'a}lez}, {Le Floc'h}, {Rieke}, {Papovich}, {Egami},
  {Alonso-Herrero}, {Engelbracht}, {Gordon}, {Misselt}, \&
  {Morrison}}]{lagache04}
{Lagache}, G., {Dole}, H., {Puget}, J.-L., {P{\'e}rez-Gonz{\'a}lez}, P.~G., {Le
  Floc'h}, E., {Rieke}, G.~H., {Papovich}, C., {Egami}, E., {Alonso-Herrero},
  A., {Engelbracht}, C.~W., {Gordon}, K.~D., {Misselt}, K.~A., \& {Morrison},
  J.~E. 2004, \apjs, 154, 112

\bibitem[{{Lagache} {et~al.}(2005){Lagache}, {Puget}, \& {Dole}}]{lagache05}
{Lagache}, G., {Puget}, J.-L., \& {Dole}, H. 2005, \araa, 43, 727

\bibitem[{{Le Floc'h} {et~al.}(2005){Le Floc'h}, {Papovich}, {Dole}, {Bell},
  {Lagache}, {Rieke}, {Egami}, {P{\'e}rez-Gonz{\'a}lez}, {Alonso-Herrero},
  {Rieke}, {Blaylock}, {Engelbracht}, {Gordon}, {Hines}, {Misselt}, {Morrison},
  \& {Mould}}]{lefloch05}
{Le Floc'h}, E., {Papovich}, C., {Dole}, H., {Bell}, E.~F., {Lagache}, G.,
  {Rieke}, G.~H., {Egami}, E., {P{\'e}rez-Gonz{\'a}lez}, P.~G.,
  {Alonso-Herrero}, A., {Rieke}, M.~J., {Blaylock}, M., {Engelbracht}, C.~W.,
  {Gordon}, K.~D., {Hines}, D.~C., {Misselt}, K.~A., {Morrison}, J.~E., \&
  {Mould}, J. 2005, \apj, 632, 169

\bibitem[{{Lewis} {et~al.}(2000){Lewis}, {Challinor}, \& {Lasenby}}]{lewis00}
{Lewis}, A., {Challinor}, A., \& {Lasenby}, A. 2000, \apj, 538, 473

\bibitem[{{Lin} {et~al.}(2009){Lin}, {Partridge}, {Pober}, {Bouchefry},
  {Burke}, {Klein}, {Coish}, \& {Huffenberger}}]{lin09}
{Lin}, Y., {Partridge}, B., {Pober}, J.~C., {Bouchefry}, K.~E., {Burke}, S.,
  {Klein}, J.~N., {Coish}, J.~W., \& {Huffenberger}, K.~M. 2009, \apj, 694, 992

\bibitem[{{Lueker} {et~al.}(2009){Lueker}, {Reichardt}, {Schaffer}, {Zahn},
  {Ade}, {Aird}, {Benson}, {Bleem}, {Carlstrom}, {Chang}, {Cho}, {Crawford},
  {Crites}, {de Haan}, {Dobbs}, {George}, {Hall}, {Halverson}, {Holder},
  {Holzapfel}, {Hrubes}, {Joy}, {Keisler}, {Knox}, {Lee}, {Leitch}, {McMahon},
  {Mehl}, {Meyer}, {Mohr}, {Montroy}, {Padin}, {Plagge}, {Pryke}, {Ruhl},
  {Shaw}, {Shirokoff}, {Spieler}, {Staniszewski}, {Stark}, {Vanderlinde},
  {Vieira}, \& {Williamson}}]{lueker09}
{Lueker}, M., {Reichardt}, C.~L., {Schaffer}, K.~K., {Zahn}, O., {Ade},
  P.~A.~R., {Aird}, K.~A., {Benson}, B.~A., {Bleem}, L.~E., {Carlstrom}, J.~E.,
  {Chang}, C.~L., {Cho}, H.~M., {Crawford}, T.~M., {Crites}, A.~T., {de Haan},
  T., {Dobbs}, M.~A., {George}, E.~M., {Hall}, N.~R., {Halverson}, N.~W.,
  {Holder}, G.~P., {Holzapfel}, W.~L., {Hrubes}, J.~D., {Joy}, M., {Keisler},
  R., {Knox}, L., {Lee}, A.~T., {Leitch}, E.~M., {McMahon}, J.~J., {Mehl}, J.,
  {Meyer}, S.~S., {Mohr}, J.~J., {Montroy}, T.~E., {Padin}, S., {Plagge}, T.,
  {Pryke}, C., {Ruhl}, J.~E., {Shaw}, L., {Shirokoff}, E., {Spieler}, H.~G.,
  {Staniszewski}, Z., {Stark}, A.~A., {Vanderlinde}, K., {Vieira}, J.~D., \&
  {Williamson}, R. 2009, submitted to \apj, arXiv:0912.4317

\bibitem[{{Magliocchetti} {et~al.}(2001){Magliocchetti}, {Moscardini},
  {Panuzzo}, {Granato}, {De Zotti}, \& {Danese}}]{magliocchetti01}
{Magliocchetti}, M., {Moscardini}, L., {Panuzzo}, P., {Granato}, G.~L., {De
  Zotti}, G., \& {Danese}, L. 2001, \mnras, 325, 1553

\bibitem[{{Marsden} {et~al.}(2009){Marsden}, {Ade}, {Bock}, {Chapin}, {Devlin},
  {Dicker}, {Griffin}, {Gundersen}, {Halpern}, {Hargrave}, {Hughes}, {Klein},
  {Mauskopf}, {Magnelli}, {Moncelsi}, {Netterfield}, {Ngo}, {Olmi}, {Pascale},
  {Patanchon}, {Rex}, {Scott}, {Semisch}, {Thomas}, {Truch}, {Tucker},
  {Tucker}, {Viero}, \& {Wiebe}}]{marsden09}
{Marsden}, G., {Ade}, P.~A.~R., {Bock}, J.~J., {Chapin}, E.~L., {Devlin},
  M.~J., {Dicker}, S.~R., {Griffin}, M., {Gundersen}, J.~O., {Halpern}, M.,
  {Hargrave}, P.~C., {Hughes}, D.~H., {Klein}, J., {Mauskopf}, P., {Magnelli},
  B., {Moncelsi}, L., {Netterfield}, C.~B., {Ngo}, H., {Olmi}, L., {Pascale},
  E., {Patanchon}, G., {Rex}, M., {Scott}, D., {Semisch}, C., {Thomas}, N.,
  {Truch}, M.~D.~P., {Tucker}, C., {Tucker}, G.~S., {Viero}, M.~P., \& {Wiebe},
  D.~V. 2009, \apj, 707, 1729

\bibitem[{{Meny} {et~al.}(2007){Meny}, {Gromov}, {Boudet}, {Bernard},
  {Paradis}, \& {Nayral}}]{meny07}
{Meny}, C., {Gromov}, V., {Boudet}, N., {Bernard}, J., {Paradis}, D., \&
  {Nayral}, C. 2007, \aap, 468, 171

\bibitem[{{Negrello} {et~al.}(2007){Negrello}, {Perrotta},
  {Gonz{\'a}lez-Nuevo}, {Silva}, {de Zotti}, {Granato}, {Baccigalupi}, \&
  {Danese}}]{negrello07}
{Negrello}, M., {Perrotta}, F., {Gonz{\'a}lez-Nuevo}, J., {Silva}, L., {de
  Zotti}, G., {Granato}, G.~L., {Baccigalupi}, C., \& {Danese}, L. 2007,
  \mnras, 377, 1557

\bibitem[{{Neugebauer} {et~al.}(1984){Neugebauer}, {Habing}, {van Duinen},
  {Aumann}, {Baud}, {Beichman}, {Beintema}, {Boggess}, {Clegg}, {de Jong},
  {Emerson}, {Gautier}, {Gillett}, {Harris}, {Hauser}, {Houck}, {Jennings},
  {Low}, {Marsden}, {Miley}, {Olnon}, {Pottasch}, {Raimond}, {Rowan-Robinson},
  {Soifer}, {Walker}, {Wesselius}, \& {Young}}]{neugebauer84}
{Neugebauer}, G., {Habing}, H.~J., {van Duinen}, R., {Aumann}, H.~H., {Baud},
  B., {Beichman}, C.~A., {Beintema}, D.~A., {Boggess}, N., {Clegg}, P.~E., {de
  Jong}, T., {Emerson}, J.~P., {Gautier}, T.~N., {Gillett}, F.~C., {Harris},
  S., {Hauser}, M.~G., {Houck}, J.~R., {Jennings}, R.~E., {Low}, F.~J.,
  {Marsden}, P.~L., {Miley}, G., {Olnon}, F.~M., {Pottasch}, S.~R., {Raimond},
  E., {Rowan-Robinson}, M., {Soifer}, B.~T., {Walker}, R.~G., {Wesselius},
  P.~R., \& {Young}, E. 1984, \apjl, 278, L1

\bibitem[{{Papovich} {et~al.}(2004){Papovich}, {Dole}, {Egami}, {Le Floc'h},
  {P{\'e}rez-Gonz{\'a}lez}, {Alonso-Herrero}, {Bai}, {Beichman}, {Blaylock},
  {Engelbracht}, {Gordon}, {Hines}, {Misselt}, {Morrison}, {Mould},
  {Muzerolle}, {Neugebauer}, {Richards}, {Rieke}, {Rieke}, {Rigby}, {Su}, \&
  {Young}}]{papovich04}
{Papovich}, C., {Dole}, H., {Egami}, E., {Le Floc'h}, E.,
  {P{\'e}rez-Gonz{\'a}lez}, P.~G., {Alonso-Herrero}, A., {Bai}, L., {Beichman},
  C.~A., {Blaylock}, M., {Engelbracht}, C.~W., {Gordon}, K.~D., {Hines}, D.~C.,
  {Misselt}, K.~A., {Morrison}, J.~E., {Mould}, J., {Muzerolle}, J.,
  {Neugebauer}, G., {Richards}, P.~L., {Rieke}, G.~H., {Rieke}, M.~J., {Rigby},
  J.~R., {Su}, K.~Y.~L., \& {Young}, E.~T. 2004, \apjs, 154, 70

\bibitem[{{Perrotta} {et~al.}(2003){Perrotta}, {Magliocchetti}, {Baccigalupi},
  {Bartelmann}, {De Zotti}, {Granato}, {Silva}, \& {Danese}}]{perrotta03}
{Perrotta}, F., {Magliocchetti}, M., {Baccigalupi}, C., {Bartelmann}, M., {De
  Zotti}, G., {Granato}, G.~L., {Silva}, L., \& {Danese}, L. 2003, \mnras, 338,
  623

\bibitem[{{Polenta} {et~al.}(2005){Polenta}, {Marinucci}, {Balbi}, {de
  Bernardis}, {Hivon}, {Masi}, {Natoli}, \& {Vittorio}}]{polenta05}
{Polenta}, G., {Marinucci}, D., {Balbi}, A., {de Bernardis}, P., {Hivon}, E.,
  {Masi}, S., {Natoli}, P., \& {Vittorio}, N. 2005, Journal of Cosmology and
  Astro-Particle Physics, 11, 1

\bibitem[{{Puget} {et~al.}(1996){Puget}, {Abergel}, {Bernard}, {Boulanger},
  {Burton}, {Desert}, \& {Hartmann}}]{puget96}
{Puget}, J.-L., {Abergel}, A., {Bernard}, J.-P., {Boulanger}, F., {Burton},
  W.~B., {Desert}, F.-X., \& {Hartmann}, D. 1996, \aap, 308, L5+

\bibitem[{{Righi} {et~al.}(2008){Righi}, {Hern{\'a}ndez-Monteagudo}, \&
  {Sunyaev}}]{righi08}
{Righi}, M., {Hern{\'a}ndez-Monteagudo}, C., \& {Sunyaev}, R.~A. 2008, \aap,
  478, 685

\bibitem[{{Santos} {et~al.}(2003){Santos}, {Cooray}, {Haiman}, {Knox}, \&
  {Ma}}]{santos03}
{Santos}, M.~G., {Cooray}, A., {Haiman}, Z., {Knox}, L., \& {Ma}, C. 2003,
  \apj, 598, 756

\bibitem[{{Scott} \& {White}(1999)}]{scott99}
{Scott}, D. \& {White}, M. 1999, \aap, 346, 1

\bibitem[{{Sehgal} {et~al.}(2010){Sehgal}, {Bode}, {Das},
  {Hernandez-Monteagudo}, {Huffenberger}, {Lin}, {Ostriker}, \&
  {Trac}}]{sehgal10}
{Sehgal}, N., {Bode}, P., {Das}, S., {Hernandez-Monteagudo}, C.,
  {Huffenberger}, K., {Lin}, Y., {Ostriker}, J.~P., \& {Trac}, H. 2010, \apj,
  709, 920

\bibitem[{{Seiffert} {et~al.}(2009){Seiffert}, {Fixsen}, {Kogut}, {Levin},
  {Limon}, {Lubin}, {Mirel}, {Singal}, {Villela}, {Wollack}, \&
  {Wuensche}}]{seiffert09}
{Seiffert}, M., {Fixsen}, D.~J., {Kogut}, A., {Levin}, S.~M., {Limon}, M.,
  {Lubin}, P.~M., {Mirel}, P., {Singal}, J., {Villela}, T., {Wollack}, E., \&
  {Wuensche}, C.~A. 2009, ArXiv e-prints

\bibitem[{{Seljak} {et~al.}(2001){Seljak}, {Burwell}, \& {Pen}}]{seljak00}
{Seljak}, U., {Burwell}, J., \& {Pen}, U. 2001, \prd, 63, 063001

\bibitem[{{Sharp} {et~al.}(2010){Sharp}, {Marrone}, {Carlstrom}, {Culverhouse},
  {Greer}, {Hawkins}, {Hennessy}, {Joy}, {Lamb}, {Leitch}, {Loh}, {Miller},
  {Mroczkowski}, {Muchovej}, {Pryke}, \& {Woody}}]{sharp10}
{Sharp}, M.~K., {Marrone}, D.~P., {Carlstrom}, J.~E., {Culverhouse}, T.,
  {Greer}, C., {Hawkins}, D., {Hennessy}, R., {Joy}, M., {Lamb}, J.~W.,
  {Leitch}, E.~M., {Loh}, M., {Miller}, A., {Mroczkowski}, T., {Muchovej}, S.,
  {Pryke}, C., \& {Woody}, D. 2010, \apj, 713, 82

\bibitem[{{Sievers} {et~al.}(2009){Sievers}, {Mason}, {Weintraub}, {Achermann},
  {Altamirano}, {Bond}, {Bronfman}, {Bustos}, {Contaldi}, {Dickinson}, {Jones},
  {May}, {Myers}, {Oyarce}, {Padin}, {Pearson}, {Pospieszalski}, {Readhead},
  {Reeves}, {Shepherd}, {Taylor}, \& {Torres}}]{sievers09}
{Sievers}, J.~L., {Mason}, B.~S., {Weintraub}, L., {Achermann}, C.,
  {Altamirano}, P., {Bond}, J.~R., {Bronfman}, L., {Bustos}, R., {Contaldi},
  C., {Dickinson}, C., {Jones}, M.~E., {May}, J., {Myers}, S.~T., {Oyarce}, N.,
  {Padin}, S., {Pearson}, T.~J., {Pospieszalski}, M., {Readhead}, A.~C.~S.,
  {Reeves}, R., {Shepherd}, M.~C., {Taylor}, A.~C., \& {Torres}, S. 2009,
  submitted to \apj, astro-ph/0901.4540

\bibitem[{{Silva} {et~al.}(1998){Silva}, {Granato}, {Bressan}, \&
  {Danese}}]{silva98}
{Silva}, L., {Granato}, G.~L., {Bressan}, A., \& {Danese}, L. 1998, /apj, 509,
  103

\bibitem[{{Smail} {et~al.}(1997){Smail}, {Ivison}, \& {Blain}}]{smail97}
{Smail}, I., {Ivison}, R.~J., \& {Blain}, A.~W. 1997, \apjl, 490, L5

\bibitem[{{Song} {et~al.}(2003){Song}, {Cooray}, {Knox}, \&
  {Zaldarriaga}}]{song03}
{Song}, Y., {Cooray}, A., {Knox}, L., \& {Zaldarriaga}, M. 2003, \apj, 590, 664

\bibitem[{Sunyaev \& Zeldovich(1980)}]{sz1980}
Sunyaev, R.~A. \& Zeldovich, Y.~B. 1980, Mon. Not. Roy. Astron. Soc., 190, 413

\bibitem[{{The Planck Collaboration}(2006)}]{planck06}
{The Planck Collaboration}. 2006, ArXiv Astrophysics e-prints

\bibitem[{{Tristram} {et~al.}(2005){Tristram}, {Mac{\'{\i}}as-P{\'e}rez},
  {Renault}, \& {Santos}}]{tristram05}
{Tristram}, M., {Mac{\'{\i}}as-P{\'e}rez}, J.~F., {Renault}, C., \& {Santos},
  D. 2005, \mnras, 358, 833

\bibitem[{{Vieira} {et~al.}(2009){Vieira}, {Crawford}, {Switzer}, {Ade},
  {Aird}, {Ashby}, {Benson}, {Bleem}, {Brodwin}, {Carlstrom}, {Chang}, {Cho},
  {Crites}, {de Haan}, {Dobbs}, {Everett}, {George}, {Gladders}, {Hall},
  {Halverson}, {High}, {Holder}, {Holzapfel}, {Hrubes}, {Joy}, {Keisler},
  {Knox}, {Lee}, {Leitch}, {Lueker}, {Marrone}, {McIntyre}, {McMahon}, {Mehl},
  {Meyer}, {Mohr}, {Montroy}, {Padin}, {Plagge}, {Pryke}, {Reichardt}, {Ruhl},
  {Schaffer}, {Shaw}, {Shirokoff}, {Spieler}, {Stalder}, {Staniszewski},
  {Stark}, {Vanderlinde}, {Walsh}, {Williamson}, {Yang}, {Zahn}, \&
  {Zenteno}}]{vieira09}
{Vieira}, J.~D., {Crawford}, T.~M., {Switzer}, E.~R., {Ade}, P.~A.~R., {Aird},
  K.~A., {Ashby}, M.~L.~N., {Benson}, B.~A., {Bleem}, L.~E., {Brodwin}, M.,
  {Carlstrom}, J.~E., {Chang}, C.~L., {Cho}, H., {Crites}, A.~T., {de Haan},
  T., {Dobbs}, M.~A., {Everett}, W., {George}, E.~M., {Gladders}, M., {Hall},
  N.~R., {Halverson}, N.~W., {High}, F.~W., {Holder}, G.~P., {Holzapfel},
  W.~L., {Hrubes}, J.~D., {Joy}, M., {Keisler}, R., {Knox}, L., {Lee}, A.~T.,
  {Leitch}, E.~M., {Lueker}, M., {Marrone}, D.~P., {McIntyre}, V., {McMahon},
  J.~J., {Mehl}, J., {Meyer}, S.~S., {Mohr}, J.~J., {Montroy}, T.~E., {Padin},
  S., {Plagge}, T., {Pryke}, C., {Reichardt}, C.~L., {Ruhl}, J.~E., {Schaffer},
  K.~K., {Shaw}, L., {Shirokoff}, E., {Spieler}, H.~G., {Stalder}, B.,
  {Staniszewski}, Z., {Stark}, A.~A., {Vanderlinde}, K., {Walsh}, W.,
  {Williamson}, R., {Yang}, Y., {Zahn}, O., \& {Zenteno}, A. 2009, submitted to
  \apj, arXiv:0912.2338

\bibitem[{{Viero} {et~al.}(2009){Viero}, {Ade}, {Bock}, {Chapin}, {Devlin},
  {Griffin}, {Gundersen}, {Halpern}, {Hargrave}, {Hughes}, {Klein},
  {MacTavish}, {Marsden}, {Martin}, {Mauskopf}, {Moncelsi}, {Negrello},
  {Netterfield}, {Olmi}, {Pascale}, {Patanchon}, {Rex}, {Scott}, {Semisch},
  {Thomas}, {Truch}, {Tucker}, {Tucker}, \& {Wiebe}}]{viero09}
{Viero}, M.~P., {Ade}, P.~A.~R., {Bock}, J.~J., {Chapin}, E.~L., {Devlin},
  M.~J., {Griffin}, M., {Gundersen}, J.~O., {Halpern}, M., {Hargrave}, P.~C.,
  {Hughes}, D.~H., {Klein}, J., {MacTavish}, C.~J., {Marsden}, G., {Martin},
  P.~G., {Mauskopf}, P., {Moncelsi}, L., {Negrello}, M., {Netterfield}, C.~B.,
  {Olmi}, L., {Pascale}, E., {Patanchon}, G., {Rex}, M., {Scott}, D.,
  {Semisch}, C., {Thomas}, N., {Truch}, M.~D.~P., {Tucker}, C., {Tucker},
  G.~S., \& {Wiebe}, D.~V. 2009, \apj, 707, 1766

\bibitem[{Zahn {et~al.}(2005)Zahn, Zaldarriaga, Hernquist, \& McQuinn}]{zahn05}
Zahn, O., Zaldarriaga, M., Hernquist, L., \& McQuinn, M. 2005, \apj, 630, 657

\end{thebibliography}

\appendix
\section{Reconstruction of spectral properties}

The spectra used in this analysis are calibrated in CMB temperature units. We can relate temperatures to flux according to
\be
\delta I_\nu  =  \frac{dB_\nu}{dT}|_{T_{CMB}}\delta T_\nu .
\ee

If we assume that the sources each have a spectral index $\alpha$ drawn from an independent normal
distribution with mean $\bar \alpha$ and variance $\sigma_\alpha^2$ then
\bea
\delta I_\nu & = & \delta I_{\nu_*} \left(\frac{\nu}{\nu_*}\right)^{\bar \alpha + \delta \alpha}  \nonumber \\
\delta I_\nu & \simeq & \delta I_{\nu_*} \left(\frac{\nu}{\nu_*}\right)^{\bar \alpha}\left[1 + \delta \alpha \ln (\nu/\nu_*)\right] .
\eea

Taking $\nu = \nu_{150}$  and $\nu_* = \nu_{220}$, we determine the Poisson power spectra from two maps:
\bea
\langle \delta I_{220}^2 \rangle & = & \langle \delta I^2 \rangle  \nonumber \\
\langle \delta I_{150}^2 \rangle & = & \langle \delta I^2 \rangle \left(\frac{\nu_{150}}{\nu_{220}}\right)^{2\bar \alpha}\left[1 + \sigma_\alpha^2 \ln ^2(\nu_{150}/\nu_{220}) \right] \nonumber \\
\langle \delta I_{150} \delta I_{220} \rangle & = & \langle \delta I^2 \rangle (\nu_{150}/\nu_{220})^{\bar \alpha} .
\eea

We can directly solve this system to get the three unknowns in terms of the three known quantities:
\bea
\langle \delta I^2 \rangle & = & \langle \delta I_{220}^2 \rangle \nonumber \\
\bar \alpha_{150-220} & = & \frac{\ln\left(\langle \delta I_{150} \delta I_{220} \rangle / \langle \delta I_{220}^2 \rangle \right)}{\ln(\nu_{150}/\nu_{220})} \nonumber \\
\sigma_\alpha^2 & = & \left[\ln(\nu_{150}/\nu_{220})\right]^{-2}\left[\frac{\langle \delta I_{150}^2 \rangle}{\langle \delta I_{220}^2 \rangle (\nu_{150}/\nu_{220})^{2\bar \alpha_{150-220}}}-1\right] .
\eea
In practice, we find a significant degeneracy between $\bar \alpha_{150-220}$ and $\sigma_\alpha^2$ in the current data and choose to place a prior on $\sigma_\alpha^2$ for the $\bar \alpha_{150-220}$ constraints in this work.

The spectral index variation only contributes to the Poisson component, not the clustered component\footnote{This is true if we assume the  spectral index scatter is uncorrelated with the large-scale number-density fluctuations.}.  To parameterize the frequency dependence of the clustered component, we introduce an effective spectral index $\bar \alpha_{150-220}^C$ and a cross-correlation coefficient $r$, such that
\bea
\langle \delta I_{220}^2 \rangle & = & \langle \delta I_{150}^2 \rangle \left(\frac{\nu_{150}}{\nu_{220}}\right)^{2\bar \alpha_{150-220}^C} \\
\langle \delta I_{150} \delta I_{220} \rangle  & = & r \sqrt{\langle \delta I_{150}^2 \rangle \langle \delta I_{220}^2 \rangle}.
\eea
The value of $r$ depends on the overlap of the window functions.  For the LDP model, $r=0.998$ at $\ell = 3000$, with a very weak $\ell$-dependence.  

In L10, the DSFG-subtracted map is proportional to $\delta T_{150} - x \delta T_{220}$.  If this
linear combination is to minimize the contribution from a component with spectral index
$\alpha_{150-220}$ then 
\be
\label{eq:xconversion}
x = \sqrt{\langle \delta T_{220}^2\rangle/\langle\delta T_{150}^2\rangle} = \frac{dB_{220}/dT}{dB_{150}/dT}\left(\frac{150}{220}\right)^{\alpha_{150-220}} = 1.21 \left(\frac{150}{220}\right)^{\alpha_{150-220}}.
\ee

The SPT bands only nominally measure power at $150\,$GHz and $220\,$GHz, and instead have a width of about 32\% and 22\%, respectively.  For purposes of estimating spectral indices, we define effective band centers ($\nu_{\rm eff}$) such that a power-law flux ($S \propto \nu^\alpha$) integrated over each band has a ratio $S_1/S_2 = (\nu_{\rm eff,1}/\nu_{{\rm eff},2})^\alpha$.  These band centers are weakly dependent on $\alpha$, but differ significantly from the nominal values of 150 and 220 GHz.  The effective band centers for $\alpha=3.5$ ($\alpha = 0$) are 154.2 (152.0) GHz  and 221.3 (219.5) GHz.  In Section~\ref{sec:evolmodels} and in the Appendix, we use $\nu_{150} = 154.2\,$GHz and $\nu_{220} = 221.3\,$GHz  to relate ratios of the 150 GHz band to 220 GHz band fluxes to spectral indices.  Similarly, for the $dB/dT$ factors in Equation~\ref{eq:xconversion} we use the band-averaged values.

\end{document}